\documentclass[11pt]{article}
\usepackage{setspace}
\usepackage{graphicx}
\usepackage{lscape}
\usepackage{pdflscape}
\usepackage{afterpage}
\usepackage{pdfpages}
\usepackage{float}
\usepackage{booktabs}
\usepackage[utf8]{inputenc}
\usepackage{indentfirst}
\usepackage{arydshln}
\usepackage{arydshln}
\usepackage{tikz}
\usepackage{lipsum}
\usepackage{pgffor}
\usepackage{dsfont}
\usepackage{amsfonts}
\usepackage{amsmath}
\allowdisplaybreaks
\usepackage{mathtools}

\usepackage{xfrac}
\usepackage{amssymb}
\usepackage{bigints}
\usepackage{graphicx}
\usepackage{url}				
\usepackage{caption}
\usepackage{subcaption} 
\usepackage{enumitem}
\usepackage{relsize,exscale}
\usepackage{multirow}
\usepackage{natbib}
\setcitestyle{authoryear, open={(},close={)}}
\usepackage[flushleft]{threeparttable}

\usepackage{datetime}

\newdateformat{monthyeardate}{%
	\monthname[\THEMONTH] \THEYEAR}

\usepackage{titlesec}
\titleformat*{\section}{\large\bfseries}
\titleformat*{\subsection}{\large\bfseries}


\newcounter{parentnumber}
\usepackage{theorem}

\usepackage[letterpaper, margin=1in]{geometry}

\usepackage{hyperref} 
\hypersetup{
	colorlinks=true, breaklinks=true, bookmarks=true,bookmarksnumbered,
	urlcolor=blue, linkcolor=blue, citecolor=blue, 
	pdftitle={}, 
	pdfauthor={\textcopyright}, 
	pdfsubject={}, 
	pdfkeywords={}, 
	pdfcreator={pdfLaTeX}, 
	pdfproducer={LaTeX with hyperref and ClassicThesis} 
}

\begin{document}

	\setstretch{1}
	\title{{\LARGE Fight like a Woman: \\ Domestic Violence and Female Judges in Brazil\thanks{We would like to thank Sophia Bhalotra, Fernanda Estevan, Mark Hoekstra, John Eric Humpries, Marcela Mello, Cormac O'Dea, Laura Schiavon, Henrik Sigstad, and Winnie van Dijk for helpful suggestions. We also thank seminar participants at Sao Paulo School of Economics - FGV, Pontificia Universidad Católica de Chile, FEN Universidad de Chile, Universidad de la República and Universidad de Montevideo.}}}

	\author{
		Helena Laneuville\thanks{Universidade de São Paulo (FEA-USP). Email: \href{mailto:laneuvillehelena@gmail.com}{laneuvillehelena@gmail.com}}  \and Vitor Possebom\thanks{Sao Paulo School of Economics - FGV. Email: \href{mailto:vitor.possebom@fgv.br}{vitor.possebom@fgv.br}. This study was financed, in part, by the São Paulo Research Foundation (FAPESP), Brasil. Process Number \#2025/04857-0.}
	}
	\date{}

	\maketitle

	\newsavebox{\tablebox} \newlength{\tableboxwidth}


	\begin{center}

		First Draft: March 2024; This Draft: \monthyeardate\today


		%
		%
		\href{https://sites.google.com/site/vitorapossebom/working-papers}{Please click here for the most recent version}

		\

		\large{\textbf{Abstract}}
	\end{center}




	{We investigate whether female judges analyze domestic violence cases differently from their male peers.} Using data from São Paulo, Brazil, between 2011 and 2019, we find that a domestic violence case assigned to a female judge is 28\% (9.7 p.p.) more likely to result in a conviction than a case assigned to a male judge with similar career characteristics. To show that this decision gap rises due to different gender perspectives about domestic violence and not because female judges are stricter than their male counterparts in all rulings, we compare it against the gender conviction-rate gap in similar types of crime. We find that this gap for domestic violence cases is larger than the same gap for other physical assault cases (8.3 p.p.). Furthermore, we {analyze two explanatory channels for this gender conviction-rate gap for domestic violence cases: gender-based differences in evidence interpretation and gender-based sentencing criteria.} We also find that female judges write longer sentences, schedule more hearings, and write more judicial documents than their male peers when analyzing domestic violence cases. Lastly, we find that the gender conviction-rate gap has no significant impact on the probability of appeals, ruling reversals, or recidivism.

	\


	\


	\newpage

	\doublespacing

	\section{Introduction}\label{SecIntro}

	{

		Judges make highly consequential decisions that profoundly affect all parties involved in a judicial process \citep{Bhuller2020,Flaherty2024}. Additionally, their decision-making process might be influenced by their individual characteristics, including gender \citep{harris_bias_2019}. Any possible gender difference in judicial decision-making is likely to be particularly relevant when analyzing domestic violence cases for two reasons. First, women are the most frequent victims of this crime \citep{united_nations_office_on_drugs_and_crime_gender-related_2023}, possibly creating closer connections between the victim and judges of a specific gender. Second, domestic violence rulings are particularly challenging because they are highly reliant on victims' testimonies, who, differently from plaintiffs of other types of crime (e.g., theft), are often traumatized due to constant abuse \citep{beecher-monas_domestic_2001}. Consequently, judges must interpret these subjective testimonies carefully, and gender-based interpretations may influence their decisions.\footnote{For instance, the \citet{humanrightswatch_criminal_1991} argues that, in the 1980s and early 1990s, Brazilian judges failed to analyze crimes of domestic violence against women in a non-discriminatory manner. \cite{Flaherty2024} defines different types of discrimination in the criminal justice system.} Contrary to this view, one may argue that judges are highly trained professionals who must follow strict law codes, implying that their individual characteristics may not matter when making decisions.

	}

	To quantitatively contrast these two views, we investigate whether female and male judges differ in their rulings of domestic violence cases in the state of São Paulo, Brazil. In particular, we document a sizeable and significant gender conviction-rate gap, suggesting that judges' individual characteristics are correlated with their decision-making process. By leveraging the text analysis of the entire sentence, we also examine two forces driving this difference: gender-based differences in evidence interpretation and gender-based sentencing criteria. At the end, we discuss the consequences of this gender conviction-rate gap beyond the Trial Court, analyzing the outcome of domestic violence cases in the Appeals Court and the recidivism behavior of these cases' defendants. We find that the Trial Judge's gender has no effect on these outcomes.

	The result on Appeals Court outcomes is surprising because conviction decisions are more frequently appealed and reversed \citep{possebom_crime_2023}. Consequently, one may have expected female judges to face more appeals and reversals than their male peers. To understand why this is not the case, we also investigate intermediate judicial outcomes: number of words per sentence, scheduled hearings, and judicial documents. We find that the averages of these outcomes are larger for female judges than for their male peers when analyzing domestic violence cases, possibly suggesting that female judges invest more resources when presiding over this type of case.

	To implement these analyses, we collect data from São Paulo, Brazil, between 2011 and 2019 and compare conviction rates by the presiding judge's gender after controlling for the judge's career characteristics and fixed effects for pairs of court district and quarter of case assignment. Case characteristics and potential outcomes are independent of the presiding judge's gender because cases are, as mandated by law, randomly assigned to judges within each court district in Brazil. Hence, the gender conviction-rate gap captures differences in decision criteria among judges of different genders.

	In this analysis, we find that a domestic violence case assigned to a female judge is 27.8\% (9.74 p.p) more likely to result in a conviction than a case assigned to a male judge with similar career characteristics. Since this decision gap may be due to differences between male and female judges regardless of their perspectives on domestic violence, we compare this gap to the gender decision gaps for two other types of crime: misdemeanors and physical assault cases. Because we focus on domestic violence crimes that do not result in death or long-term physical impairment, misdemeanors are comparable to domestic violence cases in their sentence severity according to Brazilian Law.\footnote{Being convicted of the types of domestic violence incidents in our sample results in a criminal record, a fine and/or community service instead of incarceration.} Physical assault cases are comparable to domestic violence cases because both involve some sort of physical aggression.

	We find that the gender conviction-rate gap for domestic violence cases is 2.34 p.p. and 8.34 p.p. larger than the same gap for other misdemeanor or physical assault cases. Consequently, the gender conviction-rate gap for domestic violence cases likely arises due to different perspectives about this type of crime rather than due to female judges being more punitive than their male counterparts in general.

	{For this reason, we classify this type of gap as an in-group bias according to the definition given by \cite{shayo_judicial_2011} and \cite{Jannati2025}.} We do so because victims of domestic violence are overwhelmingly female, and members of this group (female judges) act differently from members of the other group (male judges) when analyzing this type of crime specifically.

	We then investigate two possible drivers of this in-group bias: representational account and informational account \citep{boyd_untangling_2010}. The first argues that female judges act as representatives of their group and tend to protect it more intensely than male judges do. The second one argues that female judges process the information contained in domestic violence cases in a unique way. {Our evidence indicates that the informational account is a driver of the in-group bias in domestic violence cases in São Paulo, Brazil, while it suggests that the representational account is a possible driver of our estimated gender conviction-rate gap.}

	To investigate if the salience of gender identity matters (representational account), we test (i) whether female judges are more likely to mention the relationship status between the victim and the defendant, and (ii) whether the gender difference in strictness is more pronounced in {cases flagged as intimate partner violence.}\footnote{{Domestic violence cases includes not only intimate partner violence, but also violence against other relatives (e.g., kids, parents, or siblings) and household members.}} To do so, we explore each case’s full sentence and estimate whether female judges are more likely (i) to flag the case as associated with intimate partner violence and (ii) to convict a defendant
	for a domestic violence offense and simultaneously flag the case as intimate partner violence. We find that female judges are 22.1\% (12.51 p.p.) more likely than male judges to mention the relationship status between the victim and the defendant in their rulings. We also find that the gender conviction-rate gap is 34.1\% higher (9.00 p.p.) in cases in which a female judge may identify more strongly with the victim than a male judge and statistically null in cases that are not flagged as intimate partner violence offenses. These differences are {possibly} associated with the representational account of the in-group bias because the incidence and tolerance of this type of offense are strongly associated with patriarchal gender norms involving partners \citep{heise_cross-national_2015,gonzalez_gender_2020}.\footnote{As an alternative explanation for these results, we note that intimate partner violence cases may involve more severe types of violence (e.g., repeated abuse) than other types of domestic violence and female judges might react more strongly than male judges to severe types of violence regardless of any gender identity issues.}

	To test for gender differences in the interpretation of the evidence provided in domestic violence cases (informational account), we compare the gender conviction rate between cases in which the defendant was caught by the police while committing the crime and cases without this objective piece of evidence. We find that male and female judges make similar decisions when analyzing cases in which defendants were caught in the act, but differ when deciding cases that rely on more subjective evidence. Hence, female judges process the information contained in domestic violence cases differently from their male peers, as suggested by the informational account of the in-group bias.

	Lastly, we discuss the consequences of the gender conviction-rate gap beyond the Trial Court. To do so, we analyze two types of consequences of a Trial Judge's sentences. First, we examine the outcomes of domestic violence cases in the Appeals Court, analyzing whether the Trial Judge's gender affects the likelihood of appeal and the likelihood of reversal. Second, we look at the future behavior of defendants in domestic violence cases, analyzing whether the Trial Judge's gender impacts the defendants' criminal recidivism after the final sentence.

	First, the impact of stricter female judges on appeals and ruling reversals in domestic violence cases is theoretically ambiguous. On the one hand, the positive gender conviction-rate gap may lead to female judges facing a larger reversal probability because sentences by stricter judges are more frequently reversed \citep{possebom_crime_2023}. On the other hand, female judges may exert more effort (informational account) than male judges when analyzing domestic violence cases and, consequently, their rulings might have a smaller probability of being reversed. Importantly, in our empirical context, we find that these two forces balance each other and the Trial Judge's gender has no impact on the outcome of domestic violence cases in the Appeals Court. These results are present in typical court district by quarter of assignment fixed effect regressions and in outcome-based tests \citep{Knowles2001,Hoekstra2025}. Considering these findings, we conclude that the gender conviction-rate gap does not entail a higher cost for the Appeals Courts.

	To investigate whether greater effort by female judges contributes to this null effect on appeals and reversal rates, we estimate the gender gap in intermediate judicial outputs that may serve as proxies for effort. We find that, when analyzing domestic violence cases, female judges write longer sentences (20.8\% more words), schedule more hearings (5.7\% more), write more judicial orders (40.3\% more), and make more judicial decisions (29.4\% more) than their male peers. {Their longer rulings are also present when compared with their already longer sentences in misdemeanor cases, resulting in 6.9\% more words. More importantly, their } increased number of decisions is also present when compared with their already greater number of decisions in misdemeanor and physical assault cases, resulting in 11.2\% and 17.8\% more decisions in each sample.

	Second, the impact of stricter female judges on recidivism by defendants in domestic violence cases is also theoretically ambiguous. On the one hand, the positive gender conviction-rate gap may lead to female judges facing more recidivism because stricter judges tend to convict the type of defendants whose recidivism increases when punished \citep{acerenza_was_2024}. On the other hand, female judges may be able to influence defendants' behaviors in other ways, such as directly signaling that violence against women is not tolerated (representational account). Importantly, in our empirical context, we find that these two forces balance each other, and the Trial Judge's gender has no effect on four recidivism measures. These results are present in typical court district-by-quarter-of-assignment fixed-effects regressions and in outcome-based tests \citep{Knowles2001,Hoekstra2025}. Considering these results, we conclude that the gender conviction-rate gap has no negative consequences for defendants' future criminal behavior.

	This article contributes to the literature on judicial biases.\footnote{{\cite{Jannati2025} discuss and list biases in other sectors, e.g., financial markets.}} \cite{harris_bias_2019} reviews this literature, focusing on ideology, race and gender gaps. Within the studies focusing on in-group bias, many authors find evidence of judges making different decisions when they share a group identity with defendants or plaintiffs \citep{kruttschnitt_ages_2009, boyd_untangling_2010, shayo_judicial_2011, shayo_conflict_2017, Corbi2021, cai_judges_2022, chenJudgesFavorTheir2022, Cai2024,LondonoVelez2025}. Similarly to our work, \citet{kruttschnitt_ages_2009}, \cite{boyd_untangling_2010}, \citet{Corbi2021}, \cite{cai_judges_2022}, \citet{chenJudgesFavorTheir2022}, \cite{Cai2024} and \cite{LondonoVelez2025} analyze in-group gender bias in cases of uncommon crime types (rape and human trafficking) in Finland, federal appellate cases in the U.S., labor cases in Brazil, divorce cases in China, cases with female offenders in Kenya, drunk-driving cases in China, and legal abortion requests in Colombia, respectively. {Among these articles, only our work and the work of \cite{chenJudgesFavorTheir2022} and \cite{Cai2024} analyze textual details of judicial rulings beyond their final decision. Unlike these two articles, we also analyze intermediate judicial outcomes, such as the number of scheduled hearings and the total number of decisions.}

	More importantly and differently from all the previous studies, we not only document in-group bias by female judges in domestic violence cases but also analyze the consequences of this phenomenon beyond the Trial Court. We investigate the effects of the Trial Judge's gender on the case's outcomes at the Appeals Court and on the defendant's future criminal behavior.

	{
		We also contribute to the literature on how female decision-makers differ from male decision-makers. \cite{Ranganathan2020} and \cite{Alan2025} uncover that female leaders organize the working environment differently from male leaders. \cite{Chattopadhyay2004} and \cite{Clayton2014} focus on local female political leaders and their consequences for local governance. \cite{Chattopadhyay2004} find that local female executive leaders invest more in infrastructure that is directly relevant to women in villages in India. \cite{Clayton2014} find that female community councilors reduce the perceived influence of traditional leaders in villages in Lesotho. Our work differs from theirs by focusing on public officials in the judicial branch rather than decision-makers in private firms or in the executive branch of the State. Importantly, judges make decisions that are highly consequential to defendants, victims, and society as a whole, implying that understanding their decision-making process is fundamental to developing a real-world sense of how government organizations operate.
	}

	The paper is organized as follows. Section \ref{SecData} describes our dataset and institutional setting, while Section \ref{SecStategy} explains our empirical strategy. Moreover, Section \ref{SecResults} documents the gender conviction-rate gap, while Section \ref{SecUnderstanding} discusses its driving forces. Section \ref{SecConsequences} analyzes the consequences of the gender conviction-rate gap beyond the Trial Court and the existence of gender gaps in intermediate judicial outputs. Section \ref{SecConclusion} concludes.

	\section{Institutional Details, Data Description, and Summary Statistics}\label{SecData}

	{To analyze the judicial gender decision gap in domestic violence cases, we use data on criminal cases' characteristics, the prosecuted defendants, and the judges analyzing those cases. Section \ref{SecInstitutions} briefly explains important institutional details that help us to explore these data. Additionally, Section \ref{SecDataCrime} summarizes our criminal cases and defendants dataset, while Section \ref{SecDataJudges} describes our dataset of judges deciding these cases. Lastly, Section \ref{SecSummaryStats} explains which judges are included in our sample and presents summary statistics.}

	{

		\subsection{Institutional Details}\label{SecInstitutions}

		In this section, we briefly explain the institutional details behind (i) judge selection into court districts, (ii) judge assignment to criminal cases, (iii) decisions made by trial court judges, and (iv) Appeals Court work.

		First, we discuss how workers with Law degrees become judges and choose their court districts. Being a judge in Brazil is among the most competitive careers in the public sector due to high wages, prestige, and job stability. At the beginning of their careers, trial judges are selected by highly competitive public admissions exams to integrate the São Paulo State Court. As documented by \cite{helena_laneuville_breaking_2024}, approval rates ranged from 0.37\% to 2.54\% between 2006 and 2020. At the beginning of their career, state judges start as substitute judges across multiple court districts. As explained by \cite{helena_laneuville_breaking_2024}, judges can be promoted up to three times. After the first promotion, the judge presides over a small, or initial-stage, court district. Later promotions involve moving to larger districts classified as intermediate and final stages based on the number of registered voters in each district. Promotion slots must be filled according to either a merit-based or a seniority-based criterion that ranks all judges who applied to preside over a specific court district.\footnote{For details on the merit-based criterion, see \cite{helena_laneuville_breaking_2024}.}

		Regarding female representation, the judges' career is still segregated. According to \cite{helena_laneuville_breaking_2024}, women represented 38.76\% of all trial judges, 43.77\% of substitute judges, 41.50\% of initial stage judges, 45.56\% of intermediate stage judges, and 33.51\% of final stage judges between 2007 and 2022. The share of female judges among all trial judges (38.76\%) is similar to the share of female judges in our final sample (36.7\% as documented in Section \ref{SecSummaryStats}). As the female shares over different career stages suggest, female judges in our final sample tend to be less experienced and spend more time in each career stage than male judges. For this reason, controlling for judges' job tenure is crucial in our empirical strategy (Section \ref{SecSubStategy}).

		Second, we discuss how judges are assigned to criminal cases. By Brazilian Law (Decree n. 3,689/1941), a case can be allocated to judges by either connected distribution or free distribution. In the first procedure, the crime under consideration is connected to an existing criminal process. In this scenario, the current case is then attached to the previous process and analyzed by the judge associated with the older case.\footnote{For two simple examples, consider a case where a murderer killed someone and then stole a car to flee the police or a case where a drug trafficker was also caught laundering money.} In the second procedure, the crime under consideration is not connected to an existing criminal process. The new case is then allocated to a judge working in the court district (a pre-determined geographic area) where the offense occurred. If more than one judge is serving in that court district, Brazilian Law dictates that the case must be randomly assigned to any judge in the court district.

		Specifically, the random assignment of judges within court districts in the free distribution system is regulated by article n. 75 of the Criminal Process Code (Decree n. 3,689/1941), articles n. 252 and n. 548 of the Old Civil Process Code (Law n. 5,869/1973), and articles n. 285 and n. 930 of the New Civil Process Code (Law n. 13,105/2015). The 1941 law states that cases must be analyzed by judges in the court district where the crime was committed. The other two legal texts focus on how these judges are assigned. When describing the assignment of judges to cases within court districts, the 1973 law uses the word sortition (``sorteio'' in Portuguese) while the 2015 law uses the word randomization (``aleatorização'' in Portuguese).

		The practical implementation of this sortition process in São Paulo is described by two court norms: ``Provimento CSM n. 2.346/2016'' and ``Normas de Serviço da Corregedoria (Provimentos n. 50/1989 and 30/2013)''. These state court documents establish that the sortition process is an automated computer system based on random number generators. This system was designed and is supervised by the Office of the Inspector General of the São Paulo State Court, which centralizes the assignment of judges in every court district. In conversations with court clerks and judges, this centralized mechanism ensures that cases are randomly assigned to judges within court districts. Moreover, there is no suspicion by the press that the random assignment of cases within court districts is manipulated in São Paulo or in Brazil.

		Third, a trial judge may make five mutually exclusive decisions regarding the final outcome of a criminal case. To begin with, a judge may declare a case is expired due to the statute of limitations. A judge may also dismiss a case based on procedural mistakes by the police or the prosecutor. Alternatively, a judge may sign a non-prosecution agreement with the defendant, the defense attorney and the prosecutor. In this case, the defendant receives a non-incarceration punishment (e.g., fine or community service sentence) but leaves the court without a criminal record. Another option for the judge is acquitting the defendant in which case the defendant is found not guilty and leaves the court without any type of punishment or criminal record. Lastly, the judge may convict the defendant. As a result of conviction, the defendant receives a punishment (e.g., a fine or community service) and a criminal record.\footnote{If a defendant is convicted instead of signing a non-prosecution agreement, the punishment must be harsher according to Law n. 9.099/1995.}

		After the trial judge reaches a decision, the defense attorney or the district attorney may disagree with the trial judge's ruling. In this case, either attorney may file an appeal to the Appeals Court. In this scenario, the case is analyzed by a panel of three senior judges. By majority rule, this panel decides whether to reverse the trial judge's decision.\footnote{We do not consider gender differences in the decision making of Appeals Court judges because the share of female judges at this level is very small. According to \cite{helena_laneuville_breaking_2024}, only 6.45\% of Appeals Court judges are women.}
	}

	\subsection{Data on Criminal Cases}\label{SecDataCrime}

	We collect data from all criminal cases brought to the Justice Court System in the State of São Paulo, Brazil, between January 1\textsuperscript{st}, 2011, and December 3\textsuperscript{rd}, 2019. This dataset includes the date each case was assigned to a courtroom, the court district where the crime took place, the number of defendants, the defendants' names, the judge assigned to the case, the full text of the sentence, a list of the main events of the case, the primary and secondary crime types associated to the case, information on whether the case was analyzed by the Appeals Court, and the full text of the Appeals Court's ruling.

	The data also contain two pieces of information that are crucial for the construction of our sample, which we discuss in detail in Sections \ref{SecInstitutions} and \ref{SecSummaryStats}. {The first is a dummy indicating whether the case was randomly assigned to a judge within a court district and point in time, i.e., whether the case follows the free distribution system (Section \ref{SecInstitutions}). The second is the name of the courtroom, where different courtrooms may correspond to certain legal specializations. We use this information to make sure that domestic violence cases are indeed randomly assigned within a district and not systematically assigned to (or reassigned to) judges who do not preside over cases of other types of crimes (Section \ref{SecSummaryStats}).}

	We use the list of the case's main events to measure two key case characteristics. {First, we measure whether the police caught the defendant in the act of committing the crime (\textit{in flagrante delicto}), which is considered strong evidence against the defendant.} Second, we observe whether the sheriff temporarily detained the defendant in jail before the case went to the district attorney's office and was assigned a judge. Sheriffs, who must have a Law degree in Brazil, may temporarily detain a defendant if they believe that the defendant poses a significant threat to society.

	We use the defendant's names to find the number of defendants per case. Since our unit of observation is a case-defendant pair, we use the defendants' names to define the number of pairs per case.

	{We use the full-sentence text to construct our main outcome. To do this, we adapt the classification algorithm developed by \citet{possebom_crime_2023} so that it classifies domestic violence case outcomes in a more precise manner.\footnote{See Appendix \ref{AppDetailsCrimeData} for a detailed explanation of our data-construction process and algorithm's training.} In particular, our main outcome focuses exclusively on conviction decisions. This variable equals one when the judge convicts at least one of the defendants listed in the case of at least one of the charges in the case.}

	We also use the full sentence text and the list of each case's main events to construct intermediate judicial outcomes. We measure case processing time, number of words in the final ruling, number of scheduled hearings, number of judicial orders, and number of judicial decisions.

	{Moreover, we collect information on three types of outcomes that may occur after the Trial Judge's ruling. First, we measure whether the case was analyzed by the Appeals Court.} Second, using the classification algorithm proposed by \cite{possebom_crime_2023}, we analyze the full text of the Appeals Court's ruling to measure whether the Trial Judge's ruling was reversed or not.\footnote{See Appendix \ref{AppDetailsCrimeData} for a detailed explanation of our data-construction process and algorithm's training.} Third, we observe whether the defendant recidivated after their final sentence's date and the type of the new criminal offense. To measure recidivism, we check whether the defendant's name appears in any criminal case within two years after the final sentence date.\footnote{See Appendix \ref{AppDetailsCrimeData} for a detailed explanation of our data-construction process.}

	Furthermore, the district attorney's office defines each case's primary and secondary crime type prior to judge assignment. We construct three samples based on this classification.

	First, our `domestic violence' sample contains all cases with (i) domestic violence as the main subject or (ii) physical assault as the primary type and domestic violence as the secondary type. The classification ``domestic violence'' is based on the Maria da Penha Law (Law n. 11,340/2006) and increases the punishment severity of physical assault crimes associated with domestic violence.\footnote{{Related to our research question, the Maria da Penha Law has two important features. First, it defines domestic violence as a criminal offense separate from other types of physical assault cases, a key aspect for one of our identification strategies (Equation \eqref{EqDVComparable}). Second, it creates courtrooms that specialize in domestic violence. As explained in Section \ref{SecSummaryStats}, we retain only court districts without such specialized courtrooms to ensure that judges are randomly assigned to cases within those districts. For a detailed explanation of the Maria da Penha Law and an evaluation of its impacts on domestic violence in Brazil, see \cite{Ferraz2022}.}}

	Second, our `physical assault' sample contains cases whose primary type is physical assault. These cases are associated with other forms of assault, such as street fights, bar fights, or football hooliganism. We call `other physical assault' the physical assault cases that do not involve domestic violence.

	{Third, our `misdemeanors' sample contains cases that carry a maximum incarceration sentence of less than four years. This threshold is relevant because, under Brazilian Law (Law n. 9,714/1998), being convicted of these crime types results in a criminal record but not in incarceration: sentences must be converted into fines or community service. We call `other misdemeanors' the misdemeanor offenses unrelated to domestic violence.\footnote{The domestic violence cases included in our sample are regulated by Law n. 9,714/1998. Consequently, being convicted of the types of domestic violence incidents in our sample results in a criminal record, a fine and/or community service. More severe types of domestic violence may lead to incarceration, but they are not included in our sample. These cases would be classified primarily as severe physical assault, attempted murder or murder with domestic violence as a secondary crime type.} These cases include, for example, theft (36.6\% of the `other misdemeanors' included in our sample), receiving stolen property (18.0\%), drug possession intended for personal consumption (14.7\%), crimes against the administration of justice (5.9\%), threat (5.5\%), criminal misappropriation (2.8\%), crimes against intellectual property (2.7\%), offenses against reputation (2.4\%), and a variety of less common offenses (11.4\% in total).}

	Note that the sample of domestic violence cases, other physical assault cases, and other misdemeanor cases are mutually exclusive. We present summary statistics for each of these sub-samples in Section \ref{SecSummaryStats}.

	\subsection{Judges' Career and Characteristics}\label{SecDataJudges}

	We collect information on judges' careers from two public sources of information: the annual seniority list published in the São Paulo Justice Diary (2007 to 2019) and the judges' monthly productivity spreadsheet from the São Paulo Court of Justice website (2011 to 2019).

	The seniority list ranks all active judges in the court system as of December 31\textsuperscript{st} of each year by career stage and seniority. The dataset includes their starting date as magistrates and the date they were promoted to their current career stage.\footnote{{Section \ref{SecInstitutions} explains how judges are selected and progress in their careers in São Paulo.}} We combine these pooled seniority lists into a dataset about judges' careers following \citet{helena_laneuville_breaking_2024}.\footnote{This dataset allows us to track all variations of judges' names over time. Names may vary over time due to marriage, divorce, or data entry errors.} The judge-level dataset includes (i) the date they entered the justice system, (ii) the date of every promotion they have received since December 31\textsuperscript{st}, 2007, (iii) and, if they entered the justice system before 2007, the date they reached the career stage they were in at the end of 2007.\footnote{A judge may occupy four possible career stages. They start as a substitute judge, then are promoted to the initial stage, the intermediate stage, and, lastly, the final stage (Section \ref{SecInstitutions}).} We link this dataset to the criminal cases dataset by matching the judge's name. At the end, we know, for each case-defendant pair, the name of the judge presiding over the case and their years of experience, career stage, and years at the current career stage in the year the case enters the Justice system.

	The productivity data on the São Paulo Court of Justice website include the total work done by each judge in each month and court district from January 2011 to December 2019.\footnote{For example, it measures the number of hearings or disposed cases per month for each judge in each court district.} We use this information to keep track of which judges are active in each court district-month pair.\footnote{Months are the natural unit of time for keeping track of activity and work volume because salaries in Brazil are paid monthly. Furthermore, most hiring, reallocation, and termination takes effect at the beginning of each month.}

	\subsection{Sample Definition and Summary Statistics}\label{SecSummaryStats}

	Our final dataset is at the case-defendant pair level. It is crucial for our identification strategy that the gender of the judge is randomly assigned and correctly identified for each case. To enforce these two characteristics, we adopt four steps.

	{First, we restrict the dataset to cases that were randomly allocated to judges. To do so and ensure that case characteristics are uncorrelated with judges' gender, we retain only cases allocated under the free distribution system (Section \ref{SecInstitutions}).}

	Second, since we focus on domestic violence cases, we must guarantee that this type of case is randomly allocated to a judge. This assumption would be violated if a specific judge or courtroom within the court district specialized in domestic violence cases. If there is such a specialized courtroom, then domestic violence cases bypass the free distribution system and are allocated to the specialized judge. For this reason, we restrict the sample to district-quarter pairs without a courtroom specializing in domestic violence cases.

	Third, since we compare decisions made by female and male judges, we require that judges' gender be randomly assigned to cases.\footnote{{We measure judges' gender using a two-step procedure. First, we use judges' names and apply the classification algorithm proposed in the \emph{R package} \texttt{genderBR} \citep{Meireles2021}, which defines a name as typically male if men account for 90\% or more of all individuals with this name in the 2010 and 2022 Brazilian Censuses. Since a few names are not commonly associated with a specific gender, we visit the São Paulo Court System's website, check the photos of the remaining judges and manually classify their gender. The same classification procedure was used by \cite{helena_laneuville_breaking_2024}.}} In other words, the \textit{ex-ante} probability of a female judge being assigned to each case must be different from zero or one, which would not be the case in courts without judges of both genders. To enforce this restriction, we limit the sample to cases from court district-month pairs with at least one male and one female active judge in the productivity dataset.

	Lastly, we need to ensure that the assigned judge's gender is correctly determined. To do this, we focus on cases allocated to judges between January 2011 and December 2019 and use our productivity dataset to remove all observations in which the assigned judge was not active at the court district and month of the assignment, leaving us with 168 districts.\footnote{Cases are randomly allocated to courtrooms instead of directly to individual judges. If a court district is understaffed at the time of case assignment and there is no presiding judge in the assigned courtroom, the system may record the case as being attributed to the next person who takes over the courtroom, which may take place at a later point in time.}

	Using this final sample, we compute summary statistics for cases' outcomes, cases' and defendants' predetermined characteristics, and crime types. Table \ref{TabSummaryStats} displays cases' characteristics in the columns and crime types in the rows. Domestic violence cases are more frequently convicted than other physical assault cases, but are less frequently convicted than other misdemeanor cases. Lastly, defendants in domestic violence cases are more frequently caught red-handed (\emph{in flagrante delicto}) and detained in jail by the sheriff than defendants in other physical assault cases, but less so when compared against defendants in other misdemeanor cases.

	\begin{table}[!htbp]
		\caption{Cases' Characteristics by Crime Type in \%}
		\label{TabSummaryStats}
		\begin{center}
			\begin{tabular}{ccccr}
				\hline \hline
				Crime & \multirow{2}{*}{Conviction} & Caught & Defendant & \multirow{2}{*}{Obs} \\
				Type &  &  Red-Handed & in Jail &  \\
				& (1) & (2) & (3) & (4) \\
				\hline
				\multicolumn{1}{l}{DV} & 35.5 & 31.2 & 24.2 & 4,215   \\
				\multicolumn{1}{l}{Other Mis} & 43.3 & 36.7 & 30.8 & 43,998
				\\
				\multicolumn{1}{l}{Other PA} & 30.2 & 13.7 & 10.7 & 1,688   \\
				\hline
			\end{tabular}
		\end{center}
		\footnotesize{Notes: The rows of this table indicate the different samples used in this paper: domestic violence cases (DV),  misdemeanor cases unrelated to domestic violence (Other Mis), and  physical assault cases that are not domestic violence (Other PA). The first column indicates the frequency of cases whose final ruling is conviction. The conviction dummy equals one when at least one defendant is convicted of at least one charge. The next two columns show the shares of case-defendant pairs in which the defendant was caught by the police while committing the crime and was arrested by the sheriff before trial. The last column shows the size of each sample.}
	\end{table}

	We also note that there are 110 female judges and 190 male judges in our entire sample.

	\section{Empirical Strategy and Validity Tests}\label{SecStategy}

	In this section, we explain the empirical strategy used to measure the gender conviction-rate gap and indirectly test this strategy's validity.

	\subsection{Empirical Strategy}\label{SecSubStategy}

	We measure the gender conviction-rate gap for domestic violence cases by comparing the conviction probability for similar cases allocated to female and male judges. We do this by running a linear regression model on the sample of domestic violence crimes:
	\begin{equation}\label{EqDV}
		\text{Convicted}_{it}=\beta \cdot \text{Female Judge}_{it} + \eta \cdot X_{it} +  \text{Court District}_i \times \text{Quarter}_t + \varepsilon_{it}
	\end{equation}
	{where $i$ indexes case-defendant pairs and $t$ indexes quarters, where each quarter between 2011 and 2019 receives a different value of $t$. Additionally, $\text{Convicted}_{it}$ indicates whether the case associated with pair $i$ resulted in a conviction, $\text{Female Judge}_{it}$ indicates whether the case was assigned to a female judge,\footnote{Cases are randomly assigned to courtrooms, and each courtroom has one presiding judge. This variable captures the gender of the presiding judge.} $X_{it}$ are covariates including predetermined case characteristics (an indicator for whether the defendant was detained in jail by the sheriff and an indicator for whether the police caught them committing the crime) and predetermined judge characteristics (career stage, total experience, and experience in their current career stage as of the day the case was assigned), and  ``$\text{Court District}_i \times \text{Quarter}_t$'' is a full set of court district-quarter pair fixed effects, where the ``quarter'' refers to the quarter the case was assigned to a judge. Following recommendations by \cite{Chyn2025} and \cite{Goldsmith-Pinkham2025}, we cluster our standard errors at the case level since all defendants in a case are assigned to the same judge and judges in a court district-quarter cell are independently assigned to cases.\footnote{{Considering only domestic violence cases, 3,901 cases have only one defendant, 147 cases have two defendants, and 5 cases have three or more defendants. Considering only misdemeanors cases that are not domestic violence offenses, 32,633 cases have only one defendant, 3,481 cases have two defendants, and 1,111 cases have three or more defendants. Considering only physical assault cases that are not domestic violence offenses, 1,425 cases have only one defendant, 76 cases have two defendants, and 27 cases have three or more defendants.}} We implement this clustering procedure in all our regression models.}

	Our parameter of interest, $\beta$, measures whether a female judge is more likely to convict than a male judge under the assumption that the judge's gender is independent of the case's potential outcomes after controlling for district-quarter fixed effects. This exogeneity assumption is valid according to Brazilian Law, which mandates cases to be randomly allocated to judges within court districts (Section \ref{SecInstitutions}). {The fact that judges self-select into court districts is not an issue for our identification strategy because we explore the random assignment of judges conditioning on court districts and time periods. Consequently, our identification strategy leverages gender differences of judges who chose to work in the same court district during a specific time period.}

	Even under the exogeneity assumption described in the last paragraph, the $\beta$ parameter in Equation \eqref{EqDV} may capture differences between male and female judges that are unrelated to their views on domestic violence. To measure whether this conviction-rate gap is due to different gender perspectives about this type of crime, we compare this domestic violence conviction-rate gap against the gender conviction-rate gap for similar crime types. Now, our sample consists of domestic violence cases and one type of comparable criminal case. We estimate the following linear regression model:
	\begin{align}
		\text{Convicted}_{it} & = \beta_1 \cdot \text{Female Judge}_{it} + \beta_2 \cdot \text{Domestic Violence}_{it} \nonumber \\
		\label{EqDVComparable} & \hspace{20pt}+ \beta_3 \cdot \text{Female Judge}_{it} \times \text{Domestic Violence}_{it} \\
		& \hspace{20pt} + \eta \cdot X_{it} + \text{Court District}_i \times \text{Quarter}_t + \varepsilon_{it}, \nonumber
	\end{align}
	{where ``$\text{Domestic Violence}_{it}$'' indicates whether the case associated with case-defendant pair $i$ is listed as a domestic violence offense.} We use two categories of comparable crime types: (i) offenses with similar sentences (misdemeanors) and (ii) offenses of a similar nature (physical assault).

	In Equation \eqref{EqDVComparable}, our parameter of interest is $\beta_3$. This specification accounts for the extra strictness and social unacceptability of domestic violence offenses beyond that of comparable crimes (coefficient $\beta_2$) and for the possibility that female judges are stricter than male judges in general (coefficient $\beta_1$). Consequently, the $\beta_3$ coefficient captures the part of the gender conviction-rate gap that is likely due to different gender perspectives about domestic violence cases specifically. {For this reason, we interpret this coefficient as capturing a type of in-group bias according to the definition given by \cite{shayo_judicial_2011} and \cite{Jannati2025}.} We do so because victims of domestic violence are overwhelmingly female, and members of this group (female judges) may act differently from members of the other group (male judges) when analyzing this type of crime in particular.

	This interpretation for the $\beta_3$ coefficient is valid if the judge's gender is independent of the case's potential outcome after conditioning on the type of crime and district-quarter fixed effects. This exogeneity assumption is again valid under Brazilian Law, since cases are randomly assigned to judges within court districts (Section \ref{SecInstitutions}) and crime types are determined by the district attorney prior to judge assignment.

	\subsection{Validity Tests}\label{SecValidity}

	{
		To test the validity of the identifying assumptions behind Equations \eqref{EqDV} and \eqref{EqDVComparable}, we measure whether predetermined case characteristics are correlated with the gender of the assigned judge. To do so, for each of our three samples, we regress a dummy variable indicating the assigned judge's gender on predetermined case characteristics and court-district-quarter fixed effects.

		Table \ref{TabValidityTest} reports the estimated coefficient associated with each predetermined case characteristic for these three regressions. Column (1) uses the sample containing only domestic violence cases, Column (2) uses the sample containing all misdemeanors cases, and Column (3) uses the sample containing all physical assault cases. All regressions control for court district-quarter fixed effects. Standard errors, presented in parentheses, are clustered at the case ID level. Additionally, this table presents, for each regression, the F-statistic of a hypothesis test whose null imposes that all coefficients associated with predetermined case characteristics are jointly zero.

		\begin{table}[!htb]
			\caption{Covariate Balance Test}\label{TabValidityTest}
			\begin{center}{
					\begin{tabular}{l c c c }
						\hline \hline
						& \multicolumn{3}{c}{Sample:} \\ \cline{2-4}
						& DV Only & Misdemeanors & Physical Assault \\
						& (1) & (2) & (3) \\
						\hline
						Caught Red-Handed & $0.02$ & $0.00$ & $0.01$ \\
						& $(0.03)$ & $(0.01)$ & $(0.03)$  \\
						Defendant in Jail & $-0.03$ & $-0.01$ & $-0.02$ \\
						& $(0.03)$ & $(0.01)$ & $(0.03)$ \\
						DV Indicator  &  & $-0.001$ & $0.03$ \\
						&  & $(0.007)$ & $(0.02)$ \\ \hline
						\multicolumn{4}{l}{\emph{Testing the Null Hypothesis that All Coefficients are Jointly Zero}} \\
						F-Statistic & $0.70$ & $3.19$ & $2.67$ \\
						p-Value & $0.71$ & $0.36$ & $0.45$ \\ \hline
						District-Quarter FE & $\checkmark$ & $\checkmark$ & $\checkmark$ \\
						Num. obs.  & $4,215$ & $48,213$ & $5,903$ \\ \hline
						\multicolumn{4}{l}{\footnotesize{$^{***}p<0.01$; $^{**}p<0.05$; $^{*}p<0.1$}}
				\end{tabular}}
			\end{center}
			\footnotesize{Notes: This table regresses a dummy variable indicating the assigned judge's gender on predetermined case characteristics and court-district-quarter fixed effects. In addition to the estimated coefficients, this table presents, for each regression, the F-statistic from a hypothesis test whose null states that all coefficients associated with predetermined case characteristics are jointly zero. It also reports p-values connected with these three tests. Column (1) uses the sample containing only domestic violence cases. Column (2) uses the sample containing all misdemeanors cases. Column (3) uses the sample containing all physical assault cases. Standard errors, presented in parentheses, are clustered at the case ID level.}
		\end{table}

		{We find that, after including court district-quarter fixed effects, judge gender is uncorrelated with all case characteristics for all three samples. Additionally, all F-tests indicate no association between case characteristics and the assignment of female judges. Overall, the results in Table \ref{TabValidityTest} suggest that our identifying assumptions are valid, confirming that judges are randomly assigned within court districts as mandated by law.}

	}

	\section{Measuring the Gender Conviction-Rate Gap}\label{SecResults}

	We measure the gender conviction-rate gap using two empirical strategies. First, we estimate Equation \eqref{EqDV} and find that female judges are more likely to convict in domestic violence cases than equally senior men. Second, we estimate Equation \eqref{EqDVComparable} and find that misdemeanor and non-domestic violence physical assault cases present gender conviction-rate gaps that are small or nonexistent, whereas domestic violence cases present larger gaps when compared against either type of similar crime and a significantly greater gap when compared against physical assault cases.

	Table \ref{table: dv_difference} shows the differences by judge gender in the probability of conviction in domestic violence cases according to Equation \eqref{EqDV}. Column (1) regresses the conviction indicator on a female judge indicator and district-quarter fixed effects. We find that randomly selecting a woman or a man to preside does not significantly impact conviction rates, even though the point estimate is positive (3.66 p.p.) and relatively large (10.4\%) compared with the male judges' average conviction rate (35.04\%).

	\begin{table}[!htb]
		\caption{Differences by Judge Gender in the Probability of Conviction in Domestic Violence Cases}\label{table: dv_difference}
		\begin{center}
			\begin{tabular}{l c c c c c c}
				\hline
				& (1) & (2) & (3) & (4) & (5) & (6) \\
				\hline
				Female                         & $0.04$   & $0.08^{**}$  & $0.09^{***}$ & $0.10^{***}$ & $0.10^{***}$ & $0.10^{***}$ \\
				& $(0.03)$ & $(0.03)$     & $(0.03)$     & $(0.03)$     & $(0.03)$     & $(0.03)$     \\
				\hline
				Male Judge Conviction Rate                    & $0.35$   & $0.35$       & $0.35$       & $0.35$       & $0.35$       & $0.35$       \\
				District-Quarter FE & $\checkmark$      & $\checkmark$          & $\checkmark$          & $\checkmark$          & $\checkmark$          & $\checkmark$          \\
				Experience & & $\checkmark$ & $\checkmark$ & $\checkmark$ & $\checkmark$& $\checkmark$ \\
				Career Stage & & & $\checkmark$ & $\checkmark$ & $\checkmark$& $\checkmark$  \\
				Years in Stage & & & & $\checkmark$ & $\checkmark$& $\checkmark$  \\
				Caught Red-Handed & & & & & $\checkmark$& $\checkmark$  \\
				Defendant in Jail & & & & & & $\checkmark$ \\
				Num. obs.                   & $4,215$   & $4,215$       & $4,215$       & $4,215$       & $4,215$       & $4,215$       \\
				\hline
				\multicolumn{7}{l}{\footnotesize{$^{***}p<0.01$; $^{**}p<0.05$; $^{*}p<0.1$}}
			\end{tabular}
		\end{center}
		\footnotesize{Notes: This table presents the estimated results of Equation \eqref{EqDV} for six specifications that differ with respect to the included control variables. The first column shows differences by judge gender in the probability of conviction controlling for quarter-by-court district fixed effects. The second column includes the judge's years of experience as a control variable. The third column additionally includes the judge's career stage. The fourth column additionally includes the judge's years of experience in the current career stage. The fifth column additionally includes an indicator for whether the defendant was caught red-handed. The sixth column additionally includes an indicator for whether the defendant was arrested by the sheriff before the trial. Standard errors, reported in parentheses, are clustered at the case ID level.}
	\end{table}

	However, {as shown by \citet{helena_laneuville_breaking_2024} and discussed in Section \ref{SecInstitutions},} the average male judge and the average female judge have different career trajectories. We may therefore be concerned that judge tenure and experience may have an impact on conviction rates, biasing our estimates. For this reason, Columns (2)-(4) gradually include predetermined judges' career covariates as control variables.

	Column (2) includes judges' years of experience at the case's assignment date. This variable is particularly important because female judges are less experienced than their male peers. In particular, female judges have, on average, 16.03 years of experience while male judges have 16.48 years of experience.

	When we control for judges' years of experience in Column (2) of Table \ref{table: dv_difference}, we find a gender conviction rate of 7.61 p.p., which is statistically significant at the 5\% level. This increase in comparison with the estimate in Column (1) is unsurprising because judges' experience is positively correlated with conviction and female judges are less experienced than their male peers.\footnote{The estimated coefficient of years of experience in Column (2) is 0.01, indicating a positive correlation with conviction. {The combination of this positive coefficient with the negative difference in experience between female and male judges explains the increase in the coefficient from Column (1) to Column (2).}} Importantly, the estimated gender conviction rate in Column (2) is large (21.74\%) relatively to the male judges' average conviction rate.

	Columns (3) and (4) include judges' career stage and years in their current position as covariates in Equation \eqref{EqDV}. These additional controls are important because female judges stay longer in their current position and are less likely to occupy senior positions.\footnote{Female judges (i) are 18.8 p.p. less likely to be a final stage judge than their male peers and (ii) stay 1.46 years longer in their current career stage.} When we control for judges' career stage in Column (3), the estimated gender conviction rate increases to 9.16 p.p. and becomes statistically significant at 1\%. When we control for judges' years in their current position in Column (4), the estimated gender conviction rate increases to 9.94 p.p. and is statistically significant at the 1\% level. Moreover, the last estimate is relatively large (28.4\%) compared to the male judges' average conviction rate.

	Columns (5)-(6) additionally include pre-determined case characteristics as covariates in Equation \eqref{EqDV}. These control variables are indicators for whether the defendant was caught red-handed at the time of the crime and whether the defendant was arrested by the sheriff before the trial. Controlling for these predetermined covariates is not necessary for our identification strategy, because cases are randomly assigned to judges within court districts under Brazilian Law. However, adding these case characteristics to Equation \eqref{EqDV} shows that our estimated gender conviction rate gap is robust, as it barely changes between Columns (4) and (6). {For example, when we control judges' career covariates and case characteristics in Column (6), we find that randomly selecting a woman to preside over a domestic violence case increases the likelihood of conviction by 9.74 p.p. This effect is statistically significant at the 1\% level and relatively large (27.8\%) compared with the male judges' average conviction rate.} Overall, Table \ref{table: dv_difference} presents evidence supporting the existence of a gender conviction rate gap.

	Since the results in Table \ref{table: dv_difference} may capture differences between male and female judges beyond their gendered perspectives on domestic violence, we also estimate Equation \eqref{EqDVComparable} to compare the gender conviction-rate gap for domestic violence cases against this gap for similar cases. This additional comparison is relevant because, for each case, being analyzed by a female judge is a package treatment that involves different preferences with respect to conviction criteria, different technologies of evidence interpretation and, possibly, other unobservable factors correlated with gender.

	While Section \ref{SecUnderstanding} and Appendix \ref{AppModel} discuss the first two components of this package treatment, Equation \eqref{EqDVComparable} focuses on the last component. Since coefficient $\beta$ in Equation \eqref{EqDV} may capture the effect of omitted variables that are correlated with the judge’s gender and the judge’s decision (e.g., number of kids or marital status of each judge), coefficient $\beta_{3}$ in Equation \eqref{EqDVComparable} does not suffer from this issue if the effect of these omitted variables is the same across crime types. As a consequence, this coefficient captures the part of the gender conviction-rate gap that is likely due to different gender perspectives about domestic violence cases specifically.

	Table \ref{table: diff_in_diff} shows the estimated results associated with Equation \eqref{EqDVComparable}. The first column uses the sample of misdemeanors, therefore comparing crimes with similar sentence lengths to those in the domestic violence sub-sample. The last column uses the sample of physical assault cases, which are similar in nature to domestic violence cases, as they entail physical harm to another person. The regressions in both columns control for district-quarter fixed effects, judges' career covariates, and case characteristics.

	\begin{table}[!htb]
		\caption{Differences by Judge Gender in the Probability of Conviction Comparing Domestic Violence Cases against Similar Crimes}\label{table: diff_in_diff}
		\begin{center}{
				\begin{tabular}{l c c }
					\hline \hline
					Sample & Misdemeanors & Physical Assault \\
					& (1) & (2) \\
					\hline
					Female & $0.03^{***}$ & $0.00$ \\
					& $(0.01)$ & $(0.04)$ \\
					Domestic Violence (DV) & $-0.06^{***}$ & $0.06^{*}$ \\
					& $(0.01)$ & $(0.03)$ \\
					DV $\times$ Female & $0.02$ & $0.08^{*}$ \\
					& $(0.02)$ & $(0.05)$ \\
					\hline
					Men's CR (Non-DV) & $0.44$ & $0.32$ \\
					Men's CR (All) & $0.43$ & $0.34$ \\
					District-Quarter FE & $\checkmark$ & $\checkmark$ \\
					Career Controls & $\checkmark$ & $\checkmark$ \\
					Case Controls & $\checkmark$ & $\checkmark$ \\
					Num. obs. & 48,213 & 5,903 \\
					\hline
					\multicolumn{3}{l}{\footnotesize{$^{***}p<0.01$; $^{**}p<0.05$; $^{*}p<0.1$}}
			\end{tabular}}
		\end{center}
		\footnotesize{Notes: This table presents the estimated results of Equation \eqref{EqDVComparable}. The first column compares differences in male and female judge conviction rates in domestic violence cases to the gender difference in conviction rates for misdemeanors. The second column compares differences in male and female judge conviction rates in domestic violence cases to the gender difference in conviction rates for other physical assault crimes. All regressions control for District-Quarter fixed effects, judge's career covariates (years of experience, career stage, years of experience in the current career stage), and case characteristics (defendant was caught red-handed, defendant was arrested by the sheriff before the trial). CR stands for Conviction Rate and DV stands for Domestic Violence. Standard errors, presented in parentheses, are clustered at the case ID level.}
	\end{table}

	Table \ref{table: diff_in_diff} shows that the estimated coefficient on the female dummy ($\beta_{1}$) is small for the misdemeanors sample (3.20 p.p.) and close to zero for the physical assault sample (0.45 p.p.). This suggests that the gender conviction-rate gap is small or nonexistent for comparable crimes that are not classified as involving domestic violence.

	More importantly, Table \ref{table: diff_in_diff} shows that the coefficient on the interaction of judge gender and domestic violence ($\beta_{3}$) is 2.34 p.p. for the misdemeanors sample and 8.34 p.p. for the physical assault sample. The first coefficient, despite not being statistically significant, suggests that the gender conviction rate gap for domestic violence cases is 73.25\% larger than the same gap for misdemeanor cases. The interaction coefficient in the physical assault sample is statistically significant at the 10\% level {(p-value = 7.02\%)} and relatively large (25.7\%), compared with the male judges' average conviction rate for non-domestic violence offenses.\footnote{These results for the gender conviction rate specific to domestic violence ($\beta_{3}$) are robust to the inclusion of more flexible controls. For example, Table \ref{table: diff_in_diff flexible} includes the interactions of crime type with judges' career controls and case characteristics as covariates in Equation \eqref{EqDVComparable}. The estimated $\beta_{3}$ coefficient for the misdemeanor sample is similar to the one in Table \ref{table: diff_in_diff}, while the estimate for the physical assault sample increases to 8.90 p.p.} This effect is larger when we condition on defendants who were not caught red-handed (Table \ref{table:coefficients}): in the physical assault sample, the point estimate increases to 10.87 p.p. and becomes statistically significant at the 5\% level, reaching 34.96\% of the male judges' conviction rate.

	These results are also larger when we restrict our sample to cases whose final rulings were written before the Me Too Movement. In Table \ref{TabMeToo} in Appendix \ref{AppMeToo}, we re-estimate Equation \eqref{EqDVComparable} using only the cases before October 1\textsuperscript{st}, 2017, because judges may have changed their perspectives on gender after the Me Too Movement. We find that the coefficient on the interaction of judge gender and domestic violence ($\beta_{3}$) is larger in both samples, reaching 4.72 p.p. in the misdemeanor sample and 8.60 p.p. in the physical assault sample. Importantly, the $\beta_{3}$ coefficient in the misdemeanor sample becomes statistically significant at the 10\% level.

	The results in Tables \ref{table: diff_in_diff}, \ref{table:coefficients} and \ref{TabMeToo} suggest that the gender difference in conviction rates verified in Table \ref{table: dv_difference} is neither fully explained by the severity or the nature of domestic violence offenses nor by female judges being stricter than male judges in all rulings. Consequently, the interaction coefficient in these regressions captures the part of the gender conviction-rate gap that is specific to domestic violence cases and is likely due to different approaches to domestic violence cases between male and female judges.\footnote{This positive gender conviction rate gap seems to be followed by decreases in non-prosecution agreements and acquittals. This conclusion is based on Table \ref{TabOtherSentences}, which re-estimates Equations \eqref{EqDV} and \eqref{EqDVComparable} using three new outcome variables: (i) whether the case reached a non-prosecution agreement, (ii) whether the judge decided to dismiss the case, and (iii) whether the judge acquitted the defendant.} For this reason, this result suggests the existence of a type of in-group bias, {as defined by \cite{shayo_judicial_2011} and \cite{Jannati2025}.}\footnote{In Appendix \ref{AppModel}, we propose a simple economic model formalizing the definition of in-group bias.} In the next section, we explore the driving forces behind this in-group bias, while, in Section \ref{SecConsequences}, we investigate its consequences for the cases' later stages in the Justice System and the defendants' future behavior.

	\section{Understanding the Gender Conviction-Rate Gap}\label{SecUnderstanding}

	In this section, we show two important findings about the gender conviction rate gap and its driving forces.

	The first finding is suggestive evidence that the gap is explained by judges' identification with their own gender (Appendix \ref{SecSettingBar}). We find that female judges are more likely than male judges (i) to mention the relationship status between the victim and the defendant, and (ii) to convict and simultaneously flag the case as an intimate partner violence offense, while female and male judges are equally likely to convict and not flag the case as an intimate partner violence offense. These results suggest that the representational account \citep{boyd_untangling_2010} {might be} one of the explanatory forces of the in-group bias documented in Section \ref{SecResults} because violence between relationship partners is highly impacted by patriarchal social norms that make gender identity more salient \citep{heise_cross-national_2015,gonzalez_gender_2020}.\footnote{As an alternative explanation for these results, we note that intimate partner violence cases may involve more severe types of violence (e.g., repeated abuse) than other types of domestic violence and female judges might react more strongly than male judges to severe types of violence regardless of any gender identity issues.}

	The second finding, discussed in Section \ref{SecEvidence}, is that men and women respond differently to evidence presented in the same case. We show that female judges are more likely than male judges to convict in cases where evidence is more tenuous and subject to interpretation, but not when the evidence is objectively incriminating (i.e., when the defendant was caught red-handed). This result suggests that the informational account \citep{boyd_untangling_2010} is one of the explanatory forces of the in-group bias documented in Section \ref{SecResults}.

	In Appendix \ref{AppModel}, we propose a simple threshold-crossing model that formalizes the concepts of the representational and informational accounts.

	\subsection{Male and Female Judges Analyze Evidence Differently}\label{SecEvidence}

	In this section, we show that female judges are more likely than male judges to convict in domestic violence cases where evidence is more tenuous and subject to interpretation, but not when the evidence is objectively incriminating. We interpret this result as suggesting that the informational account \citep{boyd_untangling_2010} is one of the explanatory forces of the in-group bias documented in Section \ref{SecResults} because female judges process the information contained in domestic violence cases differently from their male peers.

	Domestic violence cases require complex evidentiary rulings. As discussed by \citet{beecher-monas_domestic_2001}, these cases may involve both victim testimony about the incident(s) and expert witness testimony about battered woman syndrome. Since such testimony is inherently subjective, female and male judges may interpret testimony evidence differently.

	Our dataset allows us to investigate whether the gender conviction-rate gap is stronger in cases where the evidence set is less clearly incriminating. For each case, we observe whether the defendant was arrested because they were caught in the act of committing the crime (i.e., \textit{in flagrante delicto}). These cases have more objective evidence, often including testimony from police officers who directly witnessed the offense.

	If the gender conviction-rate gap is stronger in cases where the defendant was not caught red-handed, then the gender leniency gap might exist because female and male judges interpret more subjective pieces of evidence differently (informational account). If the gender conviction-rate gap is similar across domestic violence cases regardless of the type of evidence presented, then there must be other forces driving this lenience gap.

	To measure the gender conviction-rate gap for cases with different evidence types, we re-estimate Equations \eqref{EqDV} and \eqref{EqDVComparable} using two sub-samples. The first regression restricts our sample to cases where the defendant was caught red-handed, while the second regression restricts our sample to cases where the defendant was not caught \textit{in flagrante delicto}.

	Table \ref{table:coefficients} reports regression results for the two different samples. The odd columns include cases where the defendant was caught red-handed, while the even columns include cases where this objective evidence was not present. Columns (1) and (2) report the results associated with Equation \eqref{EqDV}, while Columns (3)-(6) report the results associated with the interaction model described in Equation \eqref{EqDVComparable}.

	\begin{table}[!htb]
		\caption{Differences by Judge Gender in the Probability of Conviction conditional on the Type of Evidence}\label{table:coefficients}
		\begin{center}
			\begin{tabular}{l c c c c c c c c}
				\hline \hline
				& \multicolumn{8}{c}{Outcome: Conviction} \\ \cline{2-9}
				& \multicolumn{2}{c}{Sample:}& & \multicolumn{2}{c}{Sample:} & & \multicolumn{2}{c}{Sample:} \\
				& \multicolumn{2}{c}{Domestic Violence }& & \multicolumn{2}{c}{Misdemeanors} & & \multicolumn{2}{c}{Physical Assault } \\ \cline{2-3} \cline{5-6} \cline{8-9}
				\emph{In flagrante delicto}   & Yes & No & & Yes & No & & Yes & No  \\
				& (1) & (2) & & (3) & (4) & & (5) & (6)  \\
				\hline
				Fem & $0.03$ & $0.13^{***}$ & & $0.05^{***}$ & $0.02^{**}$ & & $0.06$ & $0.03$      \\
				& $(0.08)$ & $(0.04)$ & & $(0.01)$ & $(0.01)$ & & $(0.16)$ & $(0.05)$    \\
				DV & & & & $-0.16^{***}$ & $-0.03^{*}$ & & $0.07$& $0.08^{**}$ \\
				& & & & $(0.02)$ & $(0.01)$ & & $(0.10)$ & $(0.04)$    \\
				DV $\times$ Fem & & & & $0.03$ & $0.03$ & & $-0.03$ & $0.11^{**}$ \\
				& & & & $(0.04)$ & $(0.02)$ & & $(0.16)$ & $(0.05)$    \\
				\hline
				Men's CR (Non-DV) & - & - & & $0.56$ & $0.37$ & & $0.41$ & $0.31$        \\
				Men's CR (All) & $0.39$ & $0.33$ & & $0.54$ & $0.37$ & & $0.39$ & $0.33$      \\
				District-Quarter FE & $\checkmark$      & $\checkmark$         & & $\checkmark$           & $\checkmark$          & & $\checkmark$      & $\checkmark$         \\
				Career Controls & $\checkmark$      & $\checkmark$         & & $\checkmark$           & $\checkmark$          & & $\checkmark$      & $\checkmark$         \\
				Num. obs.  & $1,316$ & $2,899$ & & $17,459$ & $30,754$ & & $1,548$ & $4,355$      \\
				\hline
				\multicolumn{7}{l}{\footnotesize{$^{***}p<0.01$; $^{**}p<0.05$; $^{*}p<0.1$}}
			\end{tabular}
		\end{center}
		\footnotesize{Notes: This table compares the determinants of the probability of conviction. Cases are split into a sub-sample in which the defendant was caught red-handed (odd columns) and a sample in which the defendant was not caught red-handed (even columns), requiring the justice system to rely on less objective evidence. The first two columns are constructed using the sample of domestic violence cases only. Columns (3) and (4) use the sample of misdemeanors. Columns (5) and (6) use the sample of physical assault crimes. CR stands for Conviction Rate, and DV stands for Domestic Violence. All regressions control for District-Quarter fixed effects and judges' career covariates (years of experience, career stage, years of experience in the current career stage). Standard errors, presented in parentheses, are clustered at the case ID level.}
	\end{table}

	We find that a gender conviction-rate gap exists only in those cases where the defendant was not caught red-handed. We only find significant results for the female coefficient ($\beta$) in Column (2) and the interaction coefficient ($\beta_{3}$) in Column (6). These coefficients are large (39.6\% and 35.0\%, respectively) compared with the average conviction rate of male judges for cases without the \textit{in flagrante delicto} flag. Despite not being statistically significant, the interaction coefficient in Column (4) suggests that the gender conviction rate gap for domestic violence cases is 127.21\% larger than the same gap for misdemeanor cases when the defendant was not caught red-handed.

	These results suggest that female and male judges interpret less objective evidence differently, while they interpret objective evidence (cases flagged as \textit{in flagrante delicto}) in similar ways. These different interpretations of more subjective cases are consistent with the informational account being one possible driving force of the in-group bias documented in Section \ref{SecResults}.

	\section{Consequences of the Gender Conviction-Rate Gap}\label{SecConsequences}

	A gender conviction-rate gap for domestic violence cases, as found in Section \ref{SecResults}, may have consequences for the cases' later stages in the Justice System and the defendants' future behavior. In particular, defense attorneys, district attorneys, Appeals Court judges, and defendants may change their behavior when facing a female judge in a domestic violence case.

	In Section \ref{SecConsequencesAppeals}, we investigate whether stricter female judges may lead to more appeals and ruling reversals in domestic violence cases. Although we may expect positive gender appeal-rate and reversal-rate gaps due to the fact that stricter judges are more frequently reversed by the Appeals Court \citep[Figure 1]{possebom_crime_2023}, we find that female judges' rulings in domestic violence cases have similar or better outcomes in the Appeals Court than their male peers' rulings.

	This result is possibly explained by the informational account for the gender conviction-rate gap in domestic violence cases (Section \ref{SecEvidence}). Beyond interpreting subjective evidence differently, female judges may conduct the court process differently and write more complete rulings. We investigate this possibility in Section \ref{SecConsequencesAppealsMechanism} and find that female judges write longer sentences, schedule more hearings, write more orders, and make more decisions in domestic violence cases than their male peers.

	Lastly, in Section \ref{SecConsequencesRecidivism}, we investigate whether stricter female judges may lead to more recidivism by defendants in domestic violence cases. Since stricter female judges might be more likely to convict the type of defendants whose recidivism increases when punished \citep{acerenza_was_2024}, we may expect a positive gender recidivism gap. Despite this expectation, we find that female and male judges face similar recidivism probabilities in their domestic violence cases.

	\subsection{Consequences of the Gender Conviction-Rate Gap at the Appeals Court}\label{SecConsequencesAppeals}

	A gender conviction-rate gap for domestic violence cases may impact how the defense attorneys, district attorneys and Appeals Court judges treat domestic violence cases after the Trial Judge has issued a ruling. In this section, we investigate whether these agents change their behavior when facing a female judge in a domestic violence case.

	On the one hand, the phenomenon of stricter female judges for domestic violence cases may lead to more appeals and reversals of trial judges' rulings. This positive gender gap may be expected for two reasons.

	First, \citet[Figure 1 and Table J.1]{possebom_crime_2023} finds that rulings by strict judges are more frequently reversed by the Appeals Court than rulings by more lenient judges, partially because defense attorneys are more likely to appeal than district attorneys. Combining this result with our finding that female judges are stricter than male judges when analyzing domestic violence cases, we may expect a positive gender gap in the probability of domestic violence cases being analyzed by the Appeals Court and in the probability of domestic violence rulings being reversed by the Appeals Court.

	Second, \citet[Table 3]{helena_laneuville_breaking_2024} states that, between 2007 and 2022, 93.55\% of Appeals Court judges were men. If we assume that judges of both genders behave similarly at the Trial Court and the Appeals Court, our results suggest that male Appeals Court judges may be more lenient when analyzing domestic violence cases and reverse Trial Judges' conviction rulings more frequently. Consequently, we may expect a positive gender gap in the probability of domestic violence rulings being reversed by the Appeals Court.

	On the other hand, the informational account for the gender conviction-rate gap in domestic violence cases (Section \ref{SecEvidence}) may lead to a smaller probability of female judges having their rulings reversed by the Appeals Court. According to the informational account of the in-group bias, female judges analyze domestic violence evidence in a different way, possibly implying that they write more complete or careful rulings.\footnote{We investigate whether female judges write more complete rulings in Section \ref{SecConsequencesAppealsMechanism}.} When we combine this possibility with the fact that the Appeals Court focuses more on the ruling's form than the case's substance, we may expect a negative gender gap in the probability of domestic violence rulings being reversed by the Appeals Court.

	We investigate which of these two forces dominates by analyzing whether there is a gender gap in the probability of domestic violence cases being analyzed by the Appeals Court and in the probability of domestic violence rulings being reversed by the Appeals Court. To do so, we re-estimate Equations \eqref{EqDV} and \eqref{EqDVComparable} using two new outcome variables. First, our left-hand side variable is an indicator that equals 1 if there is an appeal process associated with the case and 0 otherwise. Second, our outcome variable is an indicator that equals 1 if the Appeals Court reversed the Trial Judge's ruling and 0 otherwise. To avoid sample selection issues, note that the ``zero'' group includes all cases that were not analyzed by the Appeals Court.

	Panel A of Table \ref{TabAppeals} shows the estimated coefficients of Equations \eqref{EqDV} and \eqref{EqDVComparable} when we use our appeal and reversal indicators as outcome variables. The left-hand side variable is the appeal indicator in the odd columns, while it is the reversal indicator in the even columns. The first two columns are constructed using the sample of domestic violence cases only (Equation \eqref{EqDV}). Columns (3) and (4) use the sample of misdemeanors (Equation \eqref{EqDVComparable}), while Columns (5) and (6) use the sample of physical assault crimes (Equation \eqref{EqDVComparable}). All regressions include district-quarter fixed effects and a full set of controls for case characteristics and judge career covariates.

	\begin{table}[!htbp]
		\caption{Differences by Judge Gender in the Probability of a Case Being Analyzed by the Appeals Court and of a Trial Sentence Being Reversed by the Appeals Court}\label{TabAppeals}
		\begin{center}{
				\begin{tabular}{l c c c c c c c c}
					\hline \hline
					\multicolumn{9}{c}{\emph{Panel A: All cases regardless of conviction status}}\\ \hline
					Sample: & \multicolumn{2}{c}{DV only} & & \multicolumn{2}{c}{Misdemeanors} & & \multicolumn{2}{c}{Physical Assault} \\ \cline{2-3} \cline{5-6} \cline{8-9}
					& Appealed & Reversed & & Appealed & Reversed & & Appealed & Reversed \\
					& (1) & (2) & & (3) & (4) & & (5) & (6) \\
					\hline
					Female & $0.04$ & $0.016$ & & $0.03^{***}$ & $0.001$ & & $0.07$ & $0.01$    \\
					& $(0.03)$ & $(0.012)$ & & $(0.01)$ & $(0.004)$ & & $(0.04)$ & $(0.02)$    \\
					DV &  &  & & $-0.12^{***}$ & $-0.029^{***}$ & & $0.03$ & $0.01$    \\
					&  &  & & $(0.01)$ & $(0.004)$ & & $(0.03)$ & $(0.01)$    \\
					DV x Fem &  &  & & $0.02$ & $0.01$ & & $-0.03$ & $-0.003$    \\
					&  &  & & $(0.02)$ & $(0.01)$ & & $(0.04)$ & $(0.017)$    \\
					\hline
					\multicolumn{9}{l}{\emph{Male Judges' Average Outcome for different types of cases}} \\
					Non-DV & & & & $0.36$ & $0.04$ & & $0.22$ & $0.04$    \\
					All & $0.28$ & $0.03$ & & $0.35$ & $0.04$ & & $0.26$ & $0.03$    \\ \hline
					Num. obs. & $4,215$ & $4,215$ & & $48,213$ & $48,213$ & & $5,903$ & $5,903$    \\ \hline
					\multicolumn{9}{c}{\emph{Panel B: Convicted cases only (Outcome-based Test)}} \\ \hline
					Sample: & \multicolumn{2}{c}{DV only} & & \multicolumn{2}{c}{Misdemeanors} & & \multicolumn{2}{c}{Physical Assault} \\ \cline{2-3} \cline{5-6} \cline{8-9}
					& Appealed & Reversed & & Appealed & Reversed & & Appealed & Reversed \\
					& (1) & (2) & & (3) & (4) & & (5) & (6) \\
					\hline
					Female & $-0.06$ & $0.04$ & & $0.04^{***}$ & $-0.004$ & & $0.19$ & $0.11$    \\
					& $(0.08)$ & $(0.03)$ & & $(0.01)$ & $(0.007)$ & & $(0.13)$ & $(0.07)$    \\
					DV &  &  & & $-0.13^{***}$ & $-0.05^{***}$ & & $0.09$ & $0.02$    \\
					&  &  & & $(0.02)$ & $(0.01)$ & & $(0.08)$ & $(0.03)$    \\
					DV x Fem &  &  & & $0.00$ & $0.00$ & & {$-0.22^{*}$} & $-0.09$    \\
					&  &  & & $(0.04)$ & $(0.02)$ & & $(0.13)$ & $(0.07)$    \\
					\hline
					\multicolumn{9}{l}{\emph{Male Judges' Average Outcome for different types of cases}} \\
					Non-DV & & & & $0.60$ & $0.07$ & & $0.43$ & $0.06$    \\
					All & $0.57$ & $0.05$ & & $0.60$ & $0.07$ & & $0.53$ & $0.05$    \\ \hline
					Num. obs. & $1,496$ & $1,496$ & & $20,535$ & $20,535$ & & $2,005$ & $2,005$    \\ \hline
					FE & $\checkmark$ & $\checkmark$ & & $\checkmark$ & $\checkmark$ & & $\checkmark$ & $\checkmark$ \\
					Career & $\checkmark$ & $\checkmark$ & & $\checkmark$ & $\checkmark$ & & $\checkmark$ & $\checkmark$ \\
					Case & $\checkmark$ & $\checkmark$ & & $\checkmark$ & $\checkmark$ & & $\checkmark$ & $\checkmark$ \\
					\hline
					\multicolumn{8}{l}{\footnotesize{$^{***}p<0.01$; $^{**}p<0.05$; $^{*}p<0.1$}}
			\end{tabular}}
		\end{center}
		\footnotesize{Notes: This table shows the results of Equations \eqref{EqDV} and \eqref{EqDVComparable} using two outcomes. Panel A uses all case-defendant pairs regardless of the case's conviction status. Panel B uses only case-defendant pairs whose defendants were convicted, implementing an outcome-based test. In the odd columns, the outcome indicates whether the case was analyzed by the Appeals Court, while, in the even columns, the outcome indicates whether the trial judge's sentence was reversed by the Appeals Court. The first two columns use the sample of domestic violence cases (Equation \eqref{EqDV}). Columns (3) and (4) use the sample of misdemeanors while Columns (5) and (6) use the sample of physical assault crimes (Equation \eqref{EqDVComparable}). DV stands for Domestic Violence. All regressions control for District-Quarter fixed effects (FE), judge's career covariates (years of experience, career stage, years of experience in the current career stage), and case characteristics (defendant was caught red-handed, defendant was arrested by the sheriff before the trial). Standard errors, presented in parentheses, are clustered at the case ID level.}
	\end{table}

	There are three results worth highlighting.  First, the sign and significance of the female coefficient in Columns (3) and (5) show that female judges are slightly more likely to be appealed in misdemeanor and physical assault cases, respectively. This finding is not surprising because female judges are stricter than male judges (Column (1) in Table \ref{table: diff_in_diff}), and defense attorneys appeal more frequently than district attorneys \citep[Table J.1]{possebom_crime_2023}. Second, because female coefficients are statistically null in Columns (2), (4) and (6), we find that, even though female trial judges are more frequently appealed, there is no gender difference in the reversal probability. Third and more importantly, the coefficients associated with the consequences of the in-group bias (``female'' in Columns (1) and (2) and ``DV $\times$ Fem'' in Columns (3)-(6)) are not statistically significant.\footnote{Remember that the in-group bias captures the part of the gender conviction-rate gap that is specific to domestic violence cases.} The null result in the $\beta$ coefficient in Columns (1) and (2)  means that female and male trial judges are equally likely to be appealed and reversed in domestic violence cases. The null result in the $\beta_{3}$ coefficient in Columns (3)-(6) means that the gender gap in the probability of being appealed and reversed is not significantly more intense in domestic violence cases than in other comparable cases.

	To strengthen our results on the absence of a gender gap in the appeals or reversal rates, we implement an outcome-based test in Panel B of Table \ref{TabAppeals}. This test was formalized by \cite{Knowles2001} to analyze racial gaps in vehicle searches in Maryland and is used by \cite{Hoekstra2025} to analyze racial discrimination in grand juries in Harris County, Texas.

	In our context, the outcome-based test consists of reestimating Equations \eqref{EqDV} and \eqref{EqDVComparable} using the appeal and reversal indicators as outcome variables and a subsample containing only cases with a conviction ruling at the trial phase. Intuitively, if female judges are biased against domestic violence defendants, they will convict ``less guilty'' defendants, i.e., cases with weaker evidence against the defendants. As a result, these less grounded convictions will lead to more appeals and reversals. In other words, bias against domestic violence defendants will lead to higher appeals and reversal rates in cases that were convicted by a female judge in comparison with cases that were convicted by a male judge.\footnote{{Our outcome-based test avoids the inframarginality problem typical of this method \citep{Flaherty2024} because, due to the random assignment of judges within court districts, the distributions of case complexity and evidence quality are the same for cases with female and male judges \citep{Hoekstra2025}.}} If we find that female judges' appeals and reversal rates are less than or equal to their male peers' rates when we condition on convicted cases, then we may conclude that female judges (i) are not biased against domestic violence defendants and (ii) are providing sufficiently strong arguments in their conviction rulings.

	{The estimates in Panel B of Table \ref{TabAppeals} suggest that female judges' appeals and reversal rates are similar to those of their male peers after conditioning on convicted cases. In particular, the coefficients associated with ``female'' in Columns (1) and (2) and ``DV $\times$ Fem'' in Columns (3), (4), and (6) are not statistically significant, and the coefficients in Columns (3) and (4) are very small. The only statistically significant coefficient, ``DV $\times$ Fem'' in Column (5), is negative, pointing in the opposite direction of what would be expected if female judges were biased against domestic violence defendants. Combining all these estimates, we do not reject the null hypothesis that female judges are unbiased against domestic violence defendants and are providing sufficiently strong arguments in their conviction rulings.}

	Lastly, the results in both panels of Table \ref{TabAppeals} suggest that the forces driving the gender appeal-rate and reversal-rate gaps upwards and downwards compensate each other. In other words, even though female judges are stricter than male judges when analyzing domestic violence cases, their likelihood of being appealed or reversed is less than or equal to the probabilities of their male peers. In the next section, we investigate whether differences in judges' behavior when organizing their court's work or writing their rulings may explain these results.

	\subsubsection{Mechanisms behind the Gender Conviction-Rate Gap at the Appeals Court}\label{SecConsequencesAppealsMechanism}

	Sections \ref{SecEvidence} and \ref{SecConsequencesAppeals} find that female judges interpret subjective evidence differently from their male peers and face the same appeals and reversal rate as their male peers, despite being stricter when analyzing domestic violence cases. These findings might be due to differences in how female judges conduct their court's work or in how they write their rulings.

	We investigate this possibility in this section. To do so, we analyze gender differences in case processing time in days, trial rulings' length in words, number of hearings, number of orders, and number of decisions in each case. While judicial orders include arrest warrants, witness warrants and the inclusion of new documents as evidence, judicial decisions include restraining orders, pre-trial arrest, no-money bail, and final rulings. To measure these gender differences, we re-estimate Equations \eqref{EqDV} and \eqref{EqDVComparable} using these intermediate judicial outcomes as left-hand side variables.

	Table \ref{TabIntermediateDV} shows the effects of judge gender on intermediate judicial outcomes in domestic violence cases according to Equation \eqref{EqDV}. These regressions include district-quarter fixed effects and a full set of controls for predetermined case characteristics and judge career covariates.

	\begin{table}[!htb]
		\caption{Differences by Judge Gender in Intermediate Judicial Outcomes}\label{TabIntermediateDV}
		\begin{center}
			\begin{tabular}{l c c c c c}
				\hline \hline
				& Processing & Words in & Number of & Number of & Number of \\
				& Time & Sentence & Hearings & Orders & Decisions \\
				& (1) & (2) & (3) & (4) & (5) \\
				\hline
				Female & $-35.62$ & $232.9^{***}$ & $0.127^{*}$ & $0.73^{***}$ & $0.58^{***}$ \\
				& $(26.01)$ & $(65.26)$ & $(0.069)$ & $(0.18)$ & $(0.10)$ \\
				\hline
				Men's Average & $699.2$ & $1117.5$ & $2.25$ & $1.81$ & $1.96$ \\
				District-Quarter FE & $\checkmark$ & $\checkmark$ & $\checkmark$ & $\checkmark$ & $\checkmark$  \\
				Case Controls & $\checkmark$ & $\checkmark$ & $\checkmark$ & $\checkmark$& $\checkmark$ \\
				Career Controls & $\checkmark$ & $\checkmark$ & $\checkmark$  & $\checkmark$& $\checkmark$ \\
				Num. obs. & $4,215$ & $4,215$ & $4,215$ & $4,215$ & $4,215$       \\
				\hline
				\multicolumn{6}{l}{\footnotesize{$^{***}p<0.01$; $^{**}p<0.05$; $^{*}p<0.1$}}
			\end{tabular}
		\end{center}
		\footnotesize{Notes: The regressions (Equation \eqref{EqDV}) in this table use the sample of domestic violence crimes only. The outcome in Column (1) is case processing time in days. The outcome in Column (2) is the number of words in the trial judge's ruling. The outcome in Column (3) is the number of hearings associated with the case. The outcome in Column (4) is the number of orders written by the judge. These orders include arrest warrants, witness warrants, the inclusion of new documents as evidence, and any ruling that does not involve a decision. The outcome in Column (5) is the number of decisions written by a judge. These decisions include restraining orders, pre-trial arrest, no-money bail, and final rulings. All regressions control for district-quarter fixed effects, judge's career covariates (years of experience, career stage, years of experience in the current career stage), and case characteristics (defendant was caught red-handed, defendant was arrested by the sheriff before the trial). Standard errors, presented in parentheses, are clustered at the case ID level.}
	\end{table}

	Interestingly, we find that female judges write longer sentences, schedule more hearings, write more orders, and make more decisions in domestic violence cases than their male peers (Columns (2)-(5)). The effects on the number of words, orders and decisions are statistically significant at 1\% and relatively large, representing 20.8\%, 40.3\% and 29.4\% of their male peers' average outcomes. The effect on the number of hearings is small and statistically significant only at the 10\% level. Despite these results suggesting that female judges invest more resources and effort when analyzing domestic violence cases, the small and null coefficient in Column (1) implies that female judges and male judges spend similar amounts of time analyzing domestic violence cases.

	To evaluate the robustness of these differences to omitted variables correlated with judges' gender, we estimate Equation \eqref{EqDVComparable} using intermediate judicial outcomes as left-hand side variables. The regressions in Table \ref{TabIntermediateMisdemeanorPA} include district-quarter fixed effects and a full set of predetermined controls for case characteristics and judge career covariates. Panel A reports the results using the misdemeanor sample while Panel B reports the estimates using the physical assault sample.

	\begin{table}[!htbp]
		\caption{Differences by Judge Gender in Intermediate Judicial Outcomes}\label{TabIntermediateMisdemeanorPA}
		\begin{center}{
				\begin{tabular}{l c c c c c}
					\hline \hline
					\multicolumn{6}{c}{\emph{Panel A: Misdemeanor Sample}} \\ \hline
					& Processing & Words in & Number of & Number of & Number of \\
					& Time & Sentence & Hearings & Orders & Decisions \\
					& (1) & (2) & (3) & (4) & (5) \\
					\hline
					Female & $-13.74^{*}$ & $189.1^{***}$ & $0.00$ & $0.26^{***}$ & {$0.09^{*}$} \\
					& $(7.07)$ & $(29.8)$ & $(0.02)$ & $(0.06)$ & $(0.05)$ \\
					DV & $-193.0^{***}$ & $-130.3^{***}$ & $-0.13^{***}$ & $-0.55^{***}$ & $-0.23^{***}$ \\
					& $(9.8)$ & $(33.04)$ & $(0.03)$ & $(0.06)$ & $(0.04)$ \\
					DV $\times$ Female & $10.28$ & {$88.78^{*}$} & $-0.01$ & $0.04$ & $0.25^{***}$ \\
					& $(15.83)$ & $(50.83)$ & $(0.04)$ & $(0.12)$ & $(0.07)$ \\ \hline
					\multicolumn{6}{l}{\emph{Male Judges' Average Outcome for different types of cases}} \\
					Non-DV & $773.4$ & $1296.3$ & $2.44$ & $2.03$ & $2.21$ \\
					All & $766.7$ & $1280.1$ & $2.42$ & $2.01$ & $2.19$ \\ \hline
					Num. obs. & $48,213$ & $48,213$ & $48,213$ & $48,213$ & $48,213$ \\ \hline
					\multicolumn{6}{c}{\emph{Panel B: Physical Assault Sample}} \\ \hline
					& Processing & Words in & Number of & Number of & Number of \\
					& Time & Sentence & Hearings & Orders & Decisions \\
					& (1) & (2) & (3) & (4) & (5) \\
					\hline
					Female & $24.82$ & $318.3^{***}$ & $-0.14$ & $0.67^{***}$ & $0.14$ \\
					& $(45.08)$ & $(119.3)$ & $(0.13)$ & $(0.25)$ & $(0.16)$ \\
					DV & $-146.0^{***}$ & $69.50$ & $-0.44^{***}$ & $-0.48^{***}$ & $-0.002$ \\
					& $(31.0)$ & $(68.41)$ & $(0.09)$ & $(0.15)$ & $(0.12)$ \\
					DV $\times$ Female & $-46.87$ & $-80.37$ & $0.18$ & $-0.02$ & $0.35^{**}$ \\
					& $(43.86)$ & $(122.27)$ & $(0.13)$ & $(0.27)$ & $(0.16)$ \\ \hline
					\multicolumn{6}{l}{\emph{Male Judges' Average Outcome for different types of cases}} \\
					Non-DV & $833.9$ & $1195.8$ & $2.86$ & $2.21$ & $2.06$ \\
					All & $735.8$ & $1138.8$ & $2.42$ & $1.92$ & $1.99$ \\ \hline
					Num. obs. & $5,903$ & $5,903$ & $5,903$ & $5,903$ & $5,903$ \\ \hline
					District-Quarter FE & $\checkmark$ & $\checkmark$ & $\checkmark$ & $\checkmark$ & $\checkmark$ \\
					Career Controls & $\checkmark$ & $\checkmark$ & $\checkmark$ & $\checkmark$ & $\checkmark$ \\
					Case Controls & $\checkmark$ & $\checkmark$ & $\checkmark$ & $\checkmark$ & $\checkmark$ \\
					\hline
					\multicolumn{6}{l}{\footnotesize{$^{***}p<0.01$; $^{**}p<0.05$; $^{*}p<0.1$}}
			\end{tabular}}
		\end{center}
		\footnotesize{Notes: The regressions (Equation \eqref{EqDVComparable}) in this table use the Misdemeanor sample in Panel A and the Physical Assault sample in Panel B. The outcome in Column (1) is case processing time in days. The outcome in Column (2) is the number of words in the trial judge's ruling. The outcome in Column (3) is the number of hearings associated with the case. The outcome in Column (4) is the number of orders written by the judge. These orders include arrest warrants, witness warrants, the inclusion of new documents as evidence, and any ruling that does not involve a decision. The outcome in Column (5) is the number of decisions written by a judge. These decisions include restraining orders, pre-trial arrest, no-money bail, and final rulings. DV stands for Domestic Violence. All regressions control for district-quarter fixed effects, judge's career covariates (years of experience, career stage, years of experience in the current career stage), and case characteristics (defendant was caught red-handed, defendant was arrested by the sheriff before the trial). Standard errors, presented in parentheses, are clustered at the case ID level.}
	\end{table}

	{
		Unlike Table \ref{TabIntermediateDV}, the coefficients associated with ``DV $\times$ Female'' in both panels of Table \ref{TabIntermediateMisdemeanorPA} are not statistically significant for the number of hearings and the number of orders. These results suggest that gender gaps in these variables are similar across crime types.

		However, similarly to Table \ref{TabIntermediateDV}, the $\beta_{3}$ coefficient for the number of words in the final ruling is statistically significant at the 10\% level in Panel A. This estimate is relatively large, representing 6.9\% of the average sentence length of male judges. This result weakly suggests that female judges write longer rulings than their male peers in domestic violence cases specifically.

		Additionally and similarly to Table \ref{TabIntermediateDV}, the $\beta_{3}$ coefficients for the number of decisions are statistically significant at the 1\% or 5\% levels in Panels A and B, respectively. Moreover, these estimates are relatively large, representing 11.2\% and 17.8\% of the male judges' average number of decisions. These results suggest that female judges make more decisions than their male peers in domestic violence cases specifically. These decisions are a particularly important outcome because they involve judicial measures (e.g., restriction orders and pre-trial arrest) that may protect the victim while the case is being analyzed.

	}

	Combining the results from Tables \ref{TabIntermediateDV} and \ref{TabIntermediateMisdemeanorPA}, we conclude that female judges invest the same amount or more resources than their male peers when analyzing domestic violence cases.

	\subsection{Consequences of the Gender Conviction-Rate Gap for the Defendants' Recidivism Behavior}\label{SecConsequencesRecidivism}

	A gender conviction-rate gap for domestic violence cases may impact how defendants behave after their sentences. In this section, we investigate whether these agents change their behavior when facing a female judge in a domestic violence case.

	On the one hand, the phenomenon of stricter female judges for domestic violence cases may increase the probability of recidivism. This positive gender gap may be expected for two reasons.

	First, \cite{acerenza_was_2024} analyzes misdemeanor cases in São Paulo during our sample period and finds that punishment increases two-year recidivism for agents who would be punished only by strict judges (type A defendant) while it decreases two-year recidivism for agents who would be punished by most judges, including the lenient judges (type B defendant). In our context, we find that female judges are stricter than male judges (Table \ref{table: diff_in_diff}, Column (1)), especially for domestic violence cases (Table \ref{table: dv_difference} and Column (2) in Table \ref{table: diff_in_diff}). Moreover, if we assume that female judges are stricter because they have a harsher bar for conviction (instead of having more information about domestic violence cases), then we expect female judges to convict more agents who are more likely to recidivate when punished (type A defendant) while being similarly likely to convict type B defendants. Consequently, we may expect a positive gender gap in the probability of a defendant recidivating.

	Second, domestic violence victims may be encouraged by conviction decisions to accuse their aggressors if they recidivate by committing a new domestic violence offense. In other words, victims may be more willing to report a new domestic violence incident after a conviction in the first case. If this is the case, we may expect a positive gender gap in the recidivism probability because female judges are stricter than male judges when analyzing domestic violence cases (Section \ref{SecResults}). {To investigate this possibility, we also measure the differences by judge gender in the probability of domestic violence recidivism. Additionally, to avoid this report-driven increase in the recidivism probability, we analyze the difference by judge gender in the probability of committing a new crime that is not a domestic violence offense and in the probability of committing a homicide.}

	On the other hand, the representational account for the gender conviction-rate gap in domestic violence cases (Appendix \ref{SecSettingBar}) may lead to a smaller probability of recidivism. For example, female judges may be able to directly influence the behavior of the defendants {by better signaling that violence against women is not tolerated. If the typical defendant in a domestic violence case is more likely to recidivate in a similar offense than to commit other types of crime, female judges may lead to lower recidivism.} Consequently, we may expect a negative gender gap in the recidivism probability.

	We investigate which of these two forces dominates by analyzing whether there is a gender gap in the probability of a defendant recidivating within two years of the case's final sentence.\footnote{The two-year time horizon in our domestic violence sample is sufficient to observe recidivism events ``on the streets'' because the domestic violence cases in our sample do not lead to imprisonment. Due to Law n. 9,714/1998, being convicted of the types of domestic violence incidents in our sample results in a criminal record, and a fine or community service. As a consequence, the positive gender conviction-rate gap found in Section \ref{SecResults} does not lead to less recidivism mechanically through more incarceration. Importantly, more severe types of domestic violence may lead to incarceration but they are not included in our sample. These cases would be classified primarily as severe physical assault, attempted murder or murder with domestic violence as a secondary crime type.} To do so, we re-estimate Equations \eqref{EqDV} and \eqref{EqDVComparable} using four new outcome variables. First, our left-hand side variable is an indicator that equals 1 if the defendant commits any new criminal offense within two years of their final sentence and 0 otherwise. Second, our outcome variable is an indicator that equals 1 if the defendant commits any new non-domestic violence offense within two years of their final sentence and 0 otherwise. Third, our outcome variable is an indicator that equals 1 if the defendant commits any new domestic violence offense within two years of their final sentence and 0 otherwise. Lastly, our outcome variable is an indicator that equals 1 if the defendant commits a new criminal offense within two years of their final sentence and this offense is a crime against life (i.e., murder, attempted murder, manslaughter).\footnote{We included this outcome variable because homicide cases are almost always reported to the police despite being less common offenses.}

	Panel A of Table \ref{TabRecidivismDV} shows the estimated coefficients of Equation \eqref{EqDV} when we use these four measures of recidivism as outcome variables. The columns indicate the left-hand side variables. All regressions include district-quarter fixed effects and a full set of controls for predetermined case characteristics and judge career covariates. The sample sizes in these regressions are smaller than in the previous sections because we do not use the last two years of data to avoid right-censoring.

	\begin{table}[!htb]
		\caption{Differences by Judge Gender in the Probability of a Defendant Recidivating using the Domestic Violence Sample}\label{TabRecidivismDV}
		\begin{center}{
				\begin{tabular}{l c c c c}
					\hline \hline
					\multicolumn{5}{c}{\emph{Panel A: All cases regardless of conviction status}} \\ \hline
					& Any & Excluding DV & Only DV & Murder \\
					& (1) & (2) & (3) & (4) \\
					\hline
					Female & $0.02$ & $0.01$ & $0.01$ & $0.01$ \\
					& $(0.04)$ & $(0.04)$ & $(0.02)$ & $(0.02)$ \\ \hline
					Men's Average & $0.32$ & $0.25$ & $0.07$ & $0.04$ \\
					Num. obs.  & $2,665$ & $2,665$ & $2,665$ & $2,665$  \\ \hline
					\multicolumn{5}{c}{\emph{Panel B: Convicted cases only (Outcome-based Test)}} \\ \hline
					& Any & Excluding DV & Only DV & Murder \\
					& (1) & (2) & (3) & (4) \\
					\hline
					Female & $0.11$ & $0.10$ & $0.01$ & $0.02$ \\
					& $(0.12)$ & $(0.12)$ & $(0.07)$ & $(0.05)$ \\ \hline
					Men's Average & $0.35$ & $0.28$ & $0.08$ & $0.04$ \\
					Num. obs.  & $909$ & $909$ & $909$ & $909$  \\ \hline
					District-Quarter FE & $\checkmark$ & $\checkmark$ & $\checkmark$ & $\checkmark$ \\
					Career Controls & $\checkmark$ & $\checkmark$ & $\checkmark$ & $\checkmark$ \\
					Case Controls & $\checkmark$ & $\checkmark$ & $\checkmark$ & $\checkmark$ \\
					\hline
					\multicolumn{5}{l}{\footnotesize{$^{***}p<0.01$; $^{**}p<0.05$; $^{*}p<0.1$}}
			\end{tabular}}
		\end{center}
		\footnotesize{Notes: The regressions (Equation \eqref{EqDV}) in this table use the sample of domestic violence crimes only. Panel A uses all case-defendant pairs regardless of the case's conviction status. Panel B uses only case-defendant pairs whose defendants were convicted, implementing an outcome-based test. The outcome in Column (1) indicates whether the defendant recidivated within 2 years of the date of the final ruling. The outcome in Column (2) indicates whether the defendant recidivated within 2 years of the date of the final ruling and the second offense was not a domestic violence case. The outcome in Column (3) indicates whether the defendant recidivated within 2 years of the date of the final ruling and the second offense was a domestic violence case. The outcome in Column (4) indicates whether the defendant recidivated within 2 years of the date of the final ruling and the second offense was a crime against someone's life (e.g., manslaughter, murder, attempted murder). All regressions control for District-Quarter fixed effects, judge's career covariates (years of experience, career stage, years of experience in the current career stage), and case characteristics (defendant was caught red-handed, defendant was arrested by the sheriff before the trial). Standard errors, presented in parentheses, are clustered at the case ID level. The sample size in the recidivism regressions is smaller than in the other regressions because we do not use the last two years of data to avoid right-censoring problems.}
	\end{table}

	Note that all coefficients in Panel A of Table \ref{TabRecidivismDV} are close to zero and not statistically significant. These null results mean that domestic violence defendants whose cases were analyzed by female or male trial judges are equally likely to recidivate regardless of how we measure recidivism.

	To strengthen our results on the absence of a gender gap in recidivism rates, we implement an outcome-based test in Panel B of Table \ref{TabRecidivismDV}. This test was formalized by \cite{Knowles2001} to analyze racial gaps in vehicle searches in Maryland and is used by \cite{Hoekstra2025} to analyze racial discrimination in grand juries in Harris County, Texas. In our context, the outcome-based test consists of reestimating Equation \eqref{EqDV} using the recidivism indicators as outcome variables and a subsample containing only cases with a conviction ruling at the trial phase.

	The estimates in Panel B of Table \ref{TabRecidivismDV} suggest that domestic violence defendants whose cases were analyzed by female or male trial judges are equally likely to recidivate regardless of how we measure recidivism. Once more, all coefficients are not statistically significant. Combining all these estimates, we do not reject the null hypothesis that female judges' stricter rulings are not criminogenic.

	To further boost our confidence regarding the absence of a gender gap in recidivism rates, we re-estimate Equation \eqref{EqDVComparable} using our four recidivism measures as outcome variables. We also implement the outcome-based tests associated with these regressions. These estimates are reported in Tables \ref{TabRecidivismMisdemeanor} and \ref{TabRecidivismPA}. Note that almost all the coefficients associated with the consequences of the in-group bias (``DV $\times$ Female'') are not statistically significant.\footnote{Remember that the in-group bias captures the part of the gender conviction-rate gap that is specific to domestic violence cases.} Only the $\beta_{3}$ coefficient in Column (3) in Panel B of Table \ref{TabRecidivismMisdemeanor} is statistically significant at the 10\% level. Since this coefficient {implies that female judges face larger probabilities of domestic violence recidivism,} this result might be explained by victims increasing their future reporting probabilities after a female judge convicts the defendant.\footnote{{Increasing the reporting rate of domestic violence incidents has important societal benefits \citep{Carrell2012}.}}

	When we jointly analyze Tables \ref{TabRecidivismDV}, \ref{TabRecidivismMisdemeanor} and \ref{TabRecidivismPA}, we conclude that the forces driving the gender recidivism gap upwards and downwards compensate each other. In other words, even though female judges are stricter than male judges when analyzing domestic violence cases, their stricter rulings are not criminogenic.

	\section{Conclusion}\label{SecConclusion}

	{In this paper, we provide evidence that domestic violence cases assigned to female judges are significantly more likely to result in a conviction than those assigned to male judges with similar career paths. This pattern is likely due to the reasons that are specific to domestic violence cases since the gender difference in conviction rates is stronger for cases involving domestic violence than for crimes that are similar in nature (physical assault). This finding suggests the existence of a type of in-group bias {as defined by \cite{shayo_judicial_2011} and \cite{Jannati2025}.}

		Furthermore, we find suggestive evidence for two drivers of the gender conviction-rate gap in domestic violence. The first driver is that female judges are stricter than men in cases flagged as intimate partner violence but equally strict in domestic violence cases not flagged as intimate partner violence. This finding suggests that identification with one's own gender might be a relevant factor when judges analyze domestic violence cases since the perception of intimate partner violence is highly affected by patriarchal social norms, while other forms of domestic violence are not affected by gender stereotypes. This result is consistent with the representational account \citep{boyd_untangling_2010} being a possible driver of the gender conviction-rate gap.}

	The second driver stems from gender differences in how judges analyze evidence. There is no gender gap when the defendant is caught red-handed by the police (clear and objective piece of evidence), but the gap is very strong when there is more room for subjectivity in the analysis. This result is consistent with the informational account as defined by \cite{boyd_untangling_2010}.

	Connected to gender differences in evidence interpretation, we find that female judges invest more resources than their male peers when analyzing domestic violence cases. In particular, female judges write longer sentences, schedule more hearings, write more judicial orders and make more judicial decisions than their male peers when analyzing domestic violence cases.

	Lastly, we find that the gender conviction-rate gap has no impact on the probability of appeals, ruling reversals, or recidivism for female judges in domestic violence cases. Consequently, we conclude that there is no evidence that the gender conviction-rate gap affects the use of Appeals Courts' resources or the defendants' future criminal behavior.

	Combining these results, we conclude that a diverse and representative judiciary is likely relevant in the fight against domestic violence. Increasing representation in the court system might be one inexpensive way to adjudicate crimes that disproportionately affect women.


\singlespace

\bibliography{references_part2}


\pagebreak

\newpage

\pagebreak

\setcounter{table}{0}
\renewcommand\thetable{A.\arabic{table}}

\setcounter{figure}{0}
\renewcommand\thefigure{A.\arabic{figure}}

\setcounter{equation}{0}
\renewcommand\theequation{A.\arabic{equation}}

\appendix

\begin{center}
	\huge
	Supporting Information

	(Online Appendix)

\end{center}

\doublespacing
\normalsize

\section{Additional Details about the Criminal Cases Data} \label{AppDetailsCrimeData}

\setcounter{table}{0}
\renewcommand\thetable{A.\arabic{table}}

\setcounter{figure}{0}
\renewcommand\thefigure{A.\arabic{figure}}

\setcounter{equation}{0}
\renewcommand\theequation{A.\arabic{equation}}

\setcounter{theorem}{0}
\renewcommand\thetheorem{A.\arabic{theorem}}

\setcounter{proposition}{0}
\renewcommand\theproposition{A.\arabic{proposition}}

\setcounter{corollary}{0}
\renewcommand\thecorollary{A.\arabic{corollary}}

\setcounter{assumption}{0}
\renewcommand\theassumption{A.\arabic{assumption}}

\setcounter{definition}{0}
\renewcommand\thedefinition{A.\arabic{definition}}

\setcounter{Lemma}{0}
\renewcommand\theLemma{A.\arabic{Lemma}}

In this appendix, we provide a detailed explanation of how we constructed our criminal cases dataset. This data was originally constructed by \cite{possebom_crime_2023} and is also used by \cite{acerenza_was_2024}. Here, we will explain the specific crime types included in our sample, the classification algorithms used to define which defendants were punished, and the fuzzy matching algorithm used to define which defendants recidivate. Importantly, when constructing the data for this article, we update the classification algorithms used by \cite{possebom_crime_2023} to deepen the analysis of domestic violence cases.

The final dataset was created from four initial datasets.\footnote{To construct the first three datasets, we start by collecting all case identification numbers in the São Paulo Justice Diary (``Diário Oficial da Justiça do Estado de São Paulo'') and, then, search for these cases in the State Court's website. This initial step ensures that all criminal cases are included in our sample, as they must be reported in the official state daily registry.}
\begin{enumerate}
	\item CPOPG (``Consulta de Processos de Primeiro Grau''): It contains information about all criminal cases in the Justice Court System in the State of São Paulo (TJ-SP) between 2010 and 2019. Its variables are described below.
	\begin{enumerate}
		\item \texttt{id}: An unique case identifier that can link cases across all datasets from TJ-SP.

		\item \texttt{status}: The case's status defines whether the trial judge has achieved her final decision in the trial or not, i.e., whether the case is open or not.

		\item \texttt{subject}: It denotes each case's crime type.

		\item \texttt{class}: The case's class defines whether the case's objective is to analyze whether a defendant is guilty or not. For example, some criminal cases aim to start a police investigation or to arrest a person before the trial judge's sentence.

		\item \texttt{assignment}: This variable contains information about the case's starting date and whether the case was randomly assigned to a judge within the case's court district or whether the case was assigned to a judge that was already analyzing a connected case.

		\item \texttt{trialjudge}: This variable contains the trial judge's full name.

		\item \texttt{parties}: This variable contains a list with the names of all the parties involved in the case, including defendants, prosecutors, defense attorneys and public defenders.

		\item \texttt{events}: This variable contains information on all events related to the case and their dates. For example, it contains information on occurrences at the police station (e.g., defendant was caught red-handed and defendant was arrested by the sheriff before the trial), the date of each hearing, the date and text of each judicial order, and the date and text of each judicial decision.
	\end{enumerate}

	\item CJPG (``Consulta de Julgados de Primeiro Grau''): It contains information about the final decision made by a trial judge in all criminal cases in TJ-SP between 2010 and 2019. Its variables are described below.
	\begin{enumerate}
		\item \texttt{id}: An unique case identifier that can link cases across all datasets from TJ-SP.

		\item \texttt{date}: It denotes the date of the last decision made by the case's trial judge.

		\item \texttt{courtdistrict}: It provide the court district's name.

		\item \texttt{sentence}: It provides the full text of the trial judge's final decision.
	\end{enumerate}

	\item CPOSG (``Consulta de Processos de Segundo Grau''): It contains information about all appealing criminal cases in TJ-SP between 2010 and 2019. Its variables are described below.
	\begin{enumerate}
		\item \texttt{id}: An unique case identifier that can link cases across all datasets from TJ-SP.

		\item \texttt{parties}: This variable contains a list with the names of all the parties involved in the case, including defendants, prosecutors, defense attorneys and public defenders.

		\item \texttt{composition}: It contains the name of the three Appeals judges who analyzed the appealing case, including their positions within the case (judge-rapporteur, revising judge, voting judge).

		\item \texttt{decision}: It contains the Appeals Court's final ruling.

		\item \texttt{date}: It denotes the date of the last decision made by the Appeals Court.
	\end{enumerate}

	\item Public Defenders' names: This dataset is a list with full names of all public defenders in the State of São Paulo from 2010 to 2019. It was constructed directly by the Public Defender's Office after a FOIA request. The Brazilian FOIA is known as ``Lei de Acesso à Informação'' and is regulated by Law n. 12.527/11.
\end{enumerate}

Starting from the CPOPG dataset, we implement the following steps.
\begin{enumerate}
	\item We only keep cases that are currently in the Appeals Court (\texttt{status} equal to ``Em grau de recurso'' or ``Em grau de recurso $\vert$ (Tramitação prioritária)''), closed (\texttt{status} equal to ``Extinto'', ``Extinto $\vert$ (Tramitação prioritária)'' or ``Arquivado'') or whose status is empty. Those cases are already associated with a trial judge's sentence.

	\item We only keep cases that aim to analyze whether a defendant is guilty or not. In particular, we drop all the cases whose \texttt{class} is equal to ``Execução da Pena'' (sentence execution), ``Habeas Corpus Criminal'' (habeas corpus), ``Execução Provisória'' (temporary sentence execution), ``Inquérito Policial'' (police investigation), ``Procedimento Investigatório Criminal (PIC-MP)'' (police investigation), ``Auto de Prisão em Flagrante'' (pre-trial arrest), ``Pedido de Prisão Preventiva'' (pre-trial arrest), ``Medidas Protetivas de urgência (Lei Maria da Penha) Criminal'' (urgent protective acts),``Relatório de Investigações'' (police report), ``Ação Penal de Competência do Júri'' (jury action), ``Mandado de Segurança Criminal'' (judicial mandate), ``Pedido de Busca e Apreensão Criminal'' (judicial mandate), ``Representação Criminal/Notícia de Crime'' (crime notification) and ``Termo Circunstanciado'' (report).

	\item We only keep cases whose crime types are associated with sentences that must be less than four years of incarceration or are physical assault cases. In particular, we keep cases whose \texttt{subject} is equal to ``Atentado Violento ao Pudor'' (sexual assault), ``Decorrente de Violência Doméstica'' (domestic violence), ``Violência Doméstica Contra a Mulher'' (domestic violence against a woman), ``Lesão Corporal'' (physical assault), ``Contravenções Penais'' (misdemeanors), ``Furto'' (theft), ``Furto (art. 155)'' (theft), ``Furto Privilegiado'' (qualified theft), ``Furto de coisa comum'' (theft of a common good), ``Desacato'' (contempt), ``Receptação'' (receiving stolen goods), ``Ameaça'' (threat), ``Violação de direito autoral'' (copyright violation), ``Crimes contra a Propriedade Intelectual'' (crimes against intellectual property), ``Posse de Drogas para Consumo Pessoal'' (drug consumption), ``Apropriação indébita'' (undue appropriation), ``Apropriação indébita (art. 168, caput)'' (undue appropriation), ``Quadrilha ou Bando'' (criminal conspiracy), ``Desobediência'' (disobedience), ``Resistência'' (resistence), ``Fato Atípico'' (atypical fact), ``Crimes de Abuso de Autoridade'' (abuse of authority), ``Crime Culposo'' (crime without criminal intent), ``Dano Qualificado'' (qualified harm), ``Violação de domicílio'' (trespassing), ``Favorecimento real'' (illegal favoring), ``Comunicação falsa de crime ou de contravenção'' (false criminal communication), ``Destruição / Subtração / Ocultação de Cadáver'' (destruction, subtraction or concealment of a corpse), ``Difamação'' (libel) and ``Injúria'' (insult).

	\item We only keep cases that were randomly assigned to trial judges. In particular, we keep cases whose \texttt{assignment} contain the word ``Livre''. Additionally, we only keep cases in court districts with at least one female and one male judge serving during the case's quarter. This step removes 5,281 cases from the Domestic Violence dataset, 48,498 cases from the Misdemeanors dataset, and 7,968 cases from the Physical Assault dataset.

	\item We only keep cases whose starting date is after January 1\textsuperscript{st}, 2011.
\end{enumerate}

After these steps, our Domestic Violence dataset contains 4,358 cases, our Misdemeanors dataset (including the domestic violence cases) contains 44,173 cases, and our Physical Assault dataset (including the domestic violence cases) contains 6,006 cases. We, then, merged these datasets with the CJPG dataset using cases' \texttt{id} codes.

After this step, we randomly select 65 cases per year (2010-2019) for manual classification. From this sample, we select 35 cases each year to form our training sample, ensuring that at least 10 of them are physical assault cases. Moreover, we select 10 cases each year to form our overall validation sample, and 10 domestic violence cases and 10 extra physical assault cases to form our crime-type-specific validation samples.

We manually classify these cases into six categories: ``defendant died during the trial'', ``case was expired'', ``defendant is guilty'', ``defendant accepted a non-prosecution agreement'' (``transação penal'' in Portuguese), ``case was dismissed'' (``processo suspenso'' in Portuguese) and ``defendant was acquitted''. Since some sentences are missing or incomplete, we are able to manually classify only 611 sentences.

Now, we use those 611 manually classified cases to train a classification algorithm. To do so, we divide them into a training sample (315 cases), an overall validation sample (109 sentences), a domestic violence-specific validation sample (99 sentences), and a physical assault-specific validation sample (88 sentences).

First, we design an algorithm to identify which defendants died during the trial. To do so, we check whether the sentence contains any reference to the first paragraph of Article 107 from the Brazilian Criminal Code. This specific part of the Brazilian Criminal Code states that a dead defendant cannot be punished in any way. This deterministic algorithm classifies cases into the category ``defendant died during the trial'' almost perfectly. It misclassified only one sentence that had two defendants and only one of them died before the judge made a decision.

Second, we design an algorithm to identify which cases expired. To do so, we check whether the sentence contains any reference to the fourth paragraph of Article 107 from the Brazilian Criminal Code. This deterministic algorithm classifies cases into the category ``expired case'' almost perfectly. It misclassified only four sentences that included multiple crime types in one single case.

Third, we design an algorithm to identify which cases were dismissed. To do so, we check whether the sentence contains any reference to Article 89 in Law n. 9099/95. This specific law article defines the criteria for dismissing a case. This deterministic algorithm correctly classifies 98\% of the cases into the category ``case was dismissed''. The few cases that were misclassified are cases that were initially dismissed but reopened because the defendant committed a second crime.

Fourth, we design an algorithm to identify which defendants accepted a non-prosecution agreement. To do so, we check whether the sentence contains any of the following expressions: ``cumprimento da transação penal'' (non-prosecution agreement was fulfilled), ``Homologo a proposta'' (I accept the proposition), ``homologo a transação'' (I accept the non-prosecution agreement), ``HOMOLOGO O ACORDO'' (I accept the agreement), ``proposta transação'' (A non-prosecution agreement was proposed), ``transação penal'' (non-prosecution agreement), ``Acolho a proposta'' (I accept the proposal) and ``aceitação da proposta'' (proposal acceptance). Those expressions were selected because, when manually classifying the sentences, we noticed that they were strong signals of a defendant who accepted a non-prosecution agreement. This deterministic algorithm correctly classified almost all the cases into the category ``defendant accepted a non-prosecution agreement'', making only five mistakes. In the misclassified sentences, the judge mentioned that the prosecutor proposed a non-prosecution agreement, but the defendant missed the agreement's session.

Finally, we design an algorithm to classify the remaining cases into two categories: ``defendant is guilty'' and ``defendant was acquitted''. To do so, we define a bag of words that were found to be strong signals of acquittal and guilt when we manually classified the cases in our samples. This bag of words contains the following expressions: ``absolv'' (all words related to acquittal contain this expression in Portuguese), ``art. 107, inciso IV'' (the fourth paragraph of Article 107 from the Brazilian Criminal Code defines that a defendant cannot be punished if he or she is not guilty) and related expressions, ``extinta a punibilidade'' (it means that the defendant cannot be punished) and related expressions, ``improcedente'' (unfounded), ``prescrição'' (statute of limitations), ``conden'' (all words related to punishment contain this expression in Portuguese), ``pena'' (sentence), ``procedente'' (well-founded), ``cumprimento da pena'' (sentence is fulfilled) and related expressions, ``dosimetria'' (dosimetry of the penalties) and related expressions, and ``rol dos culpados'' (book of the guilty). We, then, count how many times each one of those expressions appears in each sentence, and we normalize those counts to be between 0 and 1.

Using the normalized counts, we train six algorithms using our training sample: k-Nearest Neighbors, Random Forest, L2-Regularized Logistic Regression, L1-Regularized Logistic Regression, Naive Bayes and xgboost. We, then, validate those algorithms using our validation sample and find that the k-Nearest Neighbors algorithm correctly classifies 95.3\% of the cases, the Random Forest algorithm correctly classifies 96.5\% of the cases, the L2-Regularized Logistic Regression algorithm correctly classifies 97.7\% of the cases, the L1-Regularized Logistic Regression algorithm correctly classifies 97.9\% of the cases, the Naive Bayes algorithm correctly classifies 84.9\% of the cases and the xgboost algorithm correctly classifies 91.9\% of the cases. Additionally, the L1-Regularized Logistic Regression algorithm correctly classifies 97.9\% of the domestic violence cases and 95.3\% of the physical assault cases. Given these success rates, we use the L1-Regularized Logistic Regression algorithm to classify which trial judge rulings concluded that the ``defendant is guilty'' in our full sample.

Having designed the above algorithm, we use it to define whether the trial judge reached a conviction ruling, which is our main outcome variable. First, we find which defendants died during their trials and drop them from our sample. We then use the second, third and fourth algorithms to define which cases were expired, which cases were dismissed and which cases are associated with a non-prosecution agreement. Moreover, we use the trained L1-regularized Logistic Regression algorithm to classify the remaining cases into the categories ``defendant is guilty'' and ``defendant was acquitted''. In the end, our main outcome value equals $1$ if the case was classified into the category ``defendant is guilty'' and equals $0$ otherwise.

Now, we use the CPOSG dataset to measure whether a case was analyzed by the Appeals Court and whether the Appeals Court's ruling reverses the Trial Judge's ruling.

We start by merging the CJPG dataset with the CPOSG dataset using each case's \texttt{id} code. When merging these datasets, we create an indicator variable that denotes which cases were analyzed by the Appeals Court, i.e., which cases were matched. This is the outcome variable in Table \ref{TabAppeals}'s odd columns.

We, then,  randomly select 50 cases per year for manual classification (2010-2019) and divide them into three categories: ``cases that went to the Appeals Court, but were immediately returned due to bureaucratic errors'', ``cases whose trial judge's rulings were affirmed'' and ``cases whose trial judge's rulings were reversed''.

Now, we use those 500 manually classified cases to train a classification algorithm. To do so, we divide them into a training sample (300 cases) and a validation sample (200 sentences).

First, we design an algorithm to identify which cases went to the Appeals Court but were immediately returned. To do so, we simply check whether the Appeals Court's \texttt{decision} is empty.

Finally, we design an algorithm to classify the non-empty cases into two categories: ``cases whose trial judge's sentences were affirmed'' and ``cases whose trial judge's sentences were reversed''. To do so, we define a bag of words that were found to be strong signals of sentence reversal when we manually classified the cases in our sample. This bag of words contains the following expressions: ``absolv'' (all words related to acquittal contain this expression in Portuguese), ``art. 107, inciso IV'' (the fourth paragraph of Article 107 from the Brazilian Criminal Code defines that a defendant cannot be punished if he or she is not guilty) and related expressions, ``extinta a punibilidade'' (it means that the defendant cannot be punished) and related expressions, ``prescrição'' (statute of limitations), ``negaram provimento'' (it means that the Appeals Court affirmed the trial judge's sentence) and related expressions, ``deram provimento'' (it means that the Appeals Court reversed the trial judge's sentence) and related expressions, and ``parcial provimento'' (it means that the Appeals Court reduced the penalty established by the trial judge) and related words.  We, then, count how many times each one of those expressions appears in each sentence, and we normalize those counts to be between 0 and 1.

Using the normalized counts, we train an L1-Regularized Logistic Regression using our training sample. We then validate this algorithm using our validation sample and find that it correctly classifies 96.2\% of the cases. Given this high success rate, we use the L1-Regularized Logistic Regression algorithm to classify which trial judge's rulings were reversed in our full sample. This is the outcome variable in Table \ref{TabAppeals}'s even columns.

Now, our goal is to find the defendants' names in each case. To do so, we use the variable \texttt{parties} from the CPOPG dataset and search for names listed as ``réu'', ``ré'', ``indiciado'', ``indiciada'', ``denunciado'', ``denunciada'', ``coré'', ``coréu'', ``investigado'', ``infrator'', ``acusado'', ``autordofato'', ``autoradofato'', ``averiguada'', ``averiguado'', ``infrator'', ``querelado'', ``querelada'', ``representado'', ``reqdo'' and ``reqda''.\footnote{All these expression are associated with the word ``defendant'' in Portuguese.} Moreover, we use the variable \texttt{parties} from the CPOSG dataset and search for names listed as ``apelante'', ``recorrente'', ``requerente'', ``apelado'', ``corréu'', ``recorrido'', ``apelada'', ``réu'', ``corré'' and ``querelado''.\footnote{All these expression are associated with the word ``defendant'' or ``appealing party'' in Portuguese.} Furthermore, we analyze the full sentences from the CJPG dataset and search for names listed as  ``Réu:'', ``Ré:'', ``RÉU:'', ``RÉ:'', ``Réu'', ``Ré'', ``Autor do Fato:'', ``Autora do Fato:'', ``Autor(a) do Fato:'', ``Indiciado:'', ``Indiciada:'',  ``Sentenciado:'', ``Sentenciada:'', ``Sentenciado(a):'', ``Querelado:'', ``Querelada:'', ``Averiguado:'', ``Averiguada:'',  ``Sujeito Passivo:'', ``Denunciada:'', ``Denunciado:'', ``Requerido:'' and ``Requerida:''.\footnote{All these expressions are associated with the word ``defendant'' in Portuguese.} Finally, we delete names that are not a person's name --- such as district attorney offices, public defender offices and ``unknown author'' --- or names that are listed in the Public Defenders dataset.

After finding the defendants' names, our Domestic Violence dataset contains 4,215 case-defendant pairs, our Misdemeanors dataset (including the domestic violence cases) contains 48,213 case-defendant pairs, and our Physical Assault dataset (including the domestic violence cases) contains 5,903 case-defendant pairs. Increases or decreases in sample size when compared with the datasets containing cases are due to the fact that same cases had more than one defendant and other cases did not record any defendant's name.

Lastly, we use the CPOPG dataset to measure recidivism (Section \ref{SecConsequencesRecidivism}). We focus on whether the defendant recidivated within 2 years of their case's final sentence. A defendant $i$ in a case $j$ recidivated ($Y_{ij} = 1$) if and only if defendant $i$'s full name appears in a case $\bar{j}$ whose starting date is within 2 years after case $j$'s final sentence's date. Importantly, case $\bar{j}$ can be about any type of crime, including very severe crimes, while case $j$ has to be about domestic violence, misdemeanor, or physical assault offenses.\footnote{To observe case $\bar{j}$'s defendant, we repeat the steps in the second-to-last paragraph to find the defendants' names in a dataset that contains all cases from the CPOPG dataset, including cases that are still open and cases with severe crimes. To build this dataset, we followed the steps described above but did not subset our sample based on the variables \texttt{status} and \texttt{subject}. Moreover, when subsetting our sample based on the variable \texttt{class}, we only dropped the cases whose \texttt{class} was equal to ``Execução da Pena'', ``Habeas Corpus Criminal'', ``Execução Provisória'', ``Pedido de Busca e Apreensão Criminal'' and ``Termo Circunstanciado''. At the end, this dataset contains 1,027,120 case-defendants pairs.} To match defendants' names across cases, we use the Jaro–Winkler similarity metric and define a match if the similarity between full names in two different cases is greater than or equal to 0.95.\footnote{\cite{Abramitzky2019} match full names in historical Censuses in the U.S. and Norway. They define a match between two individuals if the Jaro–Winkler similarity between their names is greater than or equal to 0.90 and if their dates of birth match exactly. Since we do not observe defendants' dates of birth, we adopt a stricter Jaro-Winkler similarity threshold to define a match in our dataset.}

Moreover, Table \ref{TabRecidivismDV} also uses three other measures of recidivism. In Column (2), we exclude from the recidivism definition any case that is a domestic violence case. Specifically, we observe the offense type of case $\bar{j}$ and do not count it as a recidivism event if it is a domestic violence offense. In Column (3), we only count the recidivism events associated with a domestic violence case. In Column (4), we only count the recidivism events associated with crimes against someone's life (i.e., manslaughter, murder, attempted murder).

Importantly, in Section \ref{SecConsequencesRecidivism} only, we delete the case-defendant pairs whose cases started in 2018 and 2019 because their recidivism variable is not properly defined due to right-censoring. Consequently, our Domestic Violence dataset contains 2,665 case-defendant pairs, our Misdemeanors dataset (including the domestic violence cases) contains 21,322 case-defendant pairs, and our Physical Assault dataset (including the domestic violence cases) contains 3,467 case-defendant pairs.

\pagebreak

\section{Modelling the Gender Conviction-Rate Gap} \label{AppModel}

\setcounter{table}{0}
\renewcommand\thetable{B.\arabic{table}}

\setcounter{figure}{0}
\renewcommand\thefigure{B.\arabic{figure}}

\setcounter{equation}{0}
\renewcommand\theequation{B.\arabic{equation}}

\setcounter{theorem}{0}
\renewcommand\thetheorem{B.\arabic{theorem}}

\setcounter{proposition}{0}
\renewcommand\theproposition{B.\arabic{proposition}}

\setcounter{corollary}{0}
\renewcommand\thecorollary{B.\arabic{corollary}}

\setcounter{assumption}{0}
\renewcommand\theassumption{B.\arabic{assumption}}

\setcounter{definition}{0}
\renewcommand\thedefinition{B.\arabic{definition}}

\setcounter{Lemma}{0}
\renewcommand\theLemma{B.\arabic{Lemma}}

{
	In this appendix, we propose a simple economic model to formalize the concepts of gender conviction-rate gap and in-group bias \citep{shayo_judicial_2011,Jannati2025}, and their driving forces, the representational and informational accounts \citep{boyd_untangling_2010}. First, to formally define the first two concepts, we adapt the potential outcome model \citep{Rubin1972} to a scenario where cases of different crime types may be treated with a female judge rather than a male judge. Second, to formally define the representational and informational accounts of the in-group bias, we adapt the threshold-crossing model \citep{Heckman2005} to a context where the latent heterogeneity is associated with the evidence of a criminal case and the propensity score function is associated with the judge's conviction criteria.

	We start by looking at criminal cases as our observational unit.\footnote{For simplicity, we assume that every case has only one defendant.} Each case is associated with a type of crime: $d = 1$ indicates that it is a domestic violence offense, and $d = 0$ indicates that it is either a misdemeanor offense or another type of physical assault case. Moreover, each case may receive two randomly allocated treatments: $f = 1$ indicates that the case was assigned to a female judge and $f = 0$ indicates that the case was assigned to a male judge. Finally, each case has a potential conviction decision, $Y\left(f, d\right)$, which depends on the gender of the presiding judge and on the type of crime.\footnote{Note that, in our context, sex is a manipulable variable because the unit of observation is the case instead of the judge. Moreover, type of crime is a predetermined covariate, because it is chosen by the prosecutor before the presiding judge is assigned.}

	We define the gender conviction-rate gap as
	\begin{equation}
		\label{EqGap}
		\mathbb{E}\left[Y\left(1,1\right) - Y\left(0,1\right)\right] \neq 0,
	\end{equation}
	i.e., the probability that a domestic violence case ends in a conviction when presided by a female judge is different from the probability that a domestic violence case ends in a conviction when presided by a male judge.

	We also define the in-group bias as
	\begin{equation}
		\label{EqInGroup}
		\left\lbrace \begin{matrix*}[l]
			\mathbb{E}\left[Y\left(1,1\right) - Y\left(0,1\right)\right] & > & 0 \\
			\mathbb{E}\left[Y\left(1,1\right) - Y\left(0,1\right)\right] & > & \mathbb{E}\left[Y\left(1,0\right) - Y\left(0,0\right)\right]
		\end{matrix*}\right.,
	\end{equation}
	i.e., the gender conviction-rate gap for domestic violence cases (i) is positive, and (ii) is larger than the gender conviction-rate gap for other types of crime.

	Next, we formally define the representational and informational account using our empirical context and a threshold-crossing model. In any criminal case, evidence is produced against the defendant. In our model, the judge sets a conviction criterion and interprets the evidence. Then, the judge convicts the defendant if the judge interprets the evidence as being sufficient to do so.

	The evidence interpretation function, $E\colon \left\lbrace 0,1 \right\rbrace^{3} \times \mathbb{R} \rightarrow \mathbb{R}$, depends on the judge's gender (a female indicator variable, $F$), the type of crime (a domestic violence indicator variable, $D$), whether there is objective and strong evidence against the defendant (an indicator variable $R$ that is equal to one if the defendant was caught red-handed when committing the crime), and all other types of evidence produced by the police and the prosecution, $V \in \mathbb{R}$. Hence, $E\left(F,D,R,V\right)$ is a random variable capturing the amount of evidence against the defendant as interpreted by the judge assigned to the case.

	The judge's conviction criterion function, $P\colon \left\lbrace 0, 1 \right\rbrace^{3} \rightarrow \mathbb{R}$, depends on the judge's gender, the type of crime and whether the crime is associated with intimate partner violence (an indicator variable $W$ that is equal to one if the victim is the wife or intimate partner of the defendant). Hence, $P\left(F,D,W\right)$ is a random variable capturing the conviction criterion of the judged assigned to the case.

	The judge convicts the defendant if the judge interprets the evidence as being sufficient to do so, i.e.,
	\begin{equation}\label{EqThreshold}
		Y = \mathbf{1}\left\lbrace P\left(F,D,W\right) \leq E\left(F,D,R,V\right) \right\rbrace,
	\end{equation} where $Y$ denotes whether the defendant is convicted or not.

	Note that Equation \eqref{EqThreshold} imposes two exclusion restrictions. While the relationship status between the victim and the defendant --- $W$ ---- only enters the judge's conviction criterion function, being caught red-handed --- $R$ --- only enters the evidence interpretation function. The first exclusion restriction is imposed because the Criminal Process Code does not include relationship status as valid evidence against the defendant, but the Criminal Law Code includes this variable as an aggravating factor when setting the sentence. The second exclusion restriction is imposed because the Criminal Process Code considers \emph{in flagrante delicto} as a piece of evidence against the defendant but not as an aggravating factor.

	Using this model, we can formally define the representational account driving the in-group bias as
	\begin{equation}
		\label{EqRepresentational}
		P\left(1,1,1\right) - P\left(0,1,1\right) < P\left(1,1,0\right) - P\left(0,1,0\right) \leq 0,
	\end{equation}
	i.e., female judges may set stricter conviction criteria than male judges for domestic violence cases and female judges are even stricter when the case is associated with intimate partner violence.

	We can also formally define the informational account driving the in-group bias as
	\begin{equation}
		\label{EqInformational}
		E\left(1,1,0,v\right) - E\left(0,1,0,v\right) > E\left(1,1,1,v\right) - E\left(0,1,1,v\right) \geq 0,
	\end{equation}
	for any $v$ in the support of $V$. When these inequalities hold, the female judge interprets the evidence more harshly than a male judge when the defendant is caught red-handed and the female judge interprets the evidence even more harshly when the defendant is not caught red-handed.

}

{

	In the main text, we use data from criminal offenses in the State of Sao Paulo, Brazil, to test the existence of each one of the four concepts defined above. In Table \ref{table: dv_difference}, we show that the gender conviction-rate gap (Equation \eqref{EqGap}) exists. Then, in Table \ref{table: diff_in_diff}, we show that in-group bias (Equation \eqref{EqInGroup}) is present in our application. Finally, Table \ref{table:Partner} shows that our data is consistent with the representational account (Equation \eqref{EqRepresentational}) while Table \ref{table:coefficients} provides evidence that our data is consistent with the informational account (Equation \eqref{EqInformational}).
}

\newpage

\section{Female Identity as a Driver for Larger Conviction Rates }\label{SecSettingBar}

\setcounter{table}{0}
\renewcommand\thetable{C.\arabic{table}}

\setcounter{figure}{0}
\renewcommand\thefigure{C.\arabic{figure}}

\setcounter{equation}{0}
\renewcommand\theequation{C.\arabic{equation}}

\setcounter{theorem}{0}
\renewcommand\thetheorem{C.\arabic{theorem}}

\setcounter{proposition}{0}
\renewcommand\theproposition{C.\arabic{proposition}}

\setcounter{corollary}{0}
\renewcommand\thecorollary{C.\arabic{corollary}}

\setcounter{assumption}{0}
\renewcommand\theassumption{C.\arabic{assumption}}

\setcounter{definition}{0}
\renewcommand\thedefinition{C.\arabic{definition}}

\setcounter{Lemma}{0}
\renewcommand\theLemma{C.\arabic{Lemma}}

In this section, we show that the gender of the judge matters more in cases where patriarchal social norms {might be} particularly relevant: intimate partner violence. We interpret this finding as indicative that a {possible} reason for the gender conviction-rate gap is that domestic violence cases make gender identity more salient to judges, {suggesting that the representational account \citep{boyd_untangling_2010} {might be} one of the explanatory forces of the in-group bias documented in Section \ref{SecResults}} since female judges tend to protect members of their own group more intensely than male judges.

We argue that domestic violence cases make gender identity more salient because sexist social norms and gender roles significantly affect the occurrence and punishment of intimate partner violence in multiple contexts. For instance, \citet{gonzalez_gender_2020} show that, among immigrant families, the incidence of intimate partner violence is positively affected by gender norms in the country of origin. \citet{heise_cross-national_2015} show that the incidence of intimate partner violence in different countries is strongly associated with patriarchal social norms, such as male authority over women, justifications for wife beating, and laws and practices that disadvantage women in property ownership. Closer to our context, \cite{Perova2023} show that Brazilian cities with greater gender equality in labor markets have a smaller rate of female homicides.\footnote{\citet{perova_womens_2017} show that police stations that specialize in domestic violence with specially trained officers significantly reduce the number of deaths of young women by intimate partner violence in metropolitan municipalities in Brazil. \cite{Golestani2024} analyze the impact of being assigned to a court specializing in domestic violence on conviction decisions, recidivism, and revictimization. \cite{GarciaHombrados2024} analyze the impact of creating courts specialized in domestic violence on incident reporting.} In terms of judicial decision-making on domestic violence, the negative stereotypes mentioned by \citet{kafka_judging_2019} and \citet{humanrightswatch_criminal_1991} concern female intimate partners.

To show that the gender of the judge matters more for the outcome of intimate partner violence cases than for that of other domestic violence offenses, we take advantage of an important feature of our data. We are able to identify the relationship between the victim and the assailant for all cases that culminate in an active decision by the presiding judge (conviction or acquittal).\footnote{Domestic violence cases include not only intimate partner violence, but also violence against other relatives (e.g., kids, parents, or siblings) and household members.} To do this, we construct an algorithm that defines a case as intimate partner violence if, when writing the court's decision, the judge uses the Portuguese word for wife, girlfriend, fiancé, ex-wife, ex-girlfriend, ex-fiancé, or similar words.\footnote{In the domestic violence sample, 59.3\% of cases are flagged as violence against partners or ex-partners. The specific words we look for are ``namorada,'' ``ex-namorada,'' ``esposa,'' ``ex-esposa,'', ``noiva,'', ``ex-noiva,'' ``amásia,'' ``ex-amásia,'' ``companheira,'' ``ex-companheira,'' ``ex-mulher,'' ``coabitação,'' ``união estável,'' ``casamento,'' ``era casado,'' and ``era casada.'' Importantly, Portuguese is a language with well-defined grammatical gender. For this reason, the listed words refer to the relationship status of female individuals only.}

We, then, implement two tests to {indirectly assess} the importance of gender identity when analyzing domestic violence cases. Our first test verifies whether female judges are more likely to mention the relationship status between the victim and the defendant. If this is the case, female judges might be more aware of intimate partner violence than male judges, suggesting the importance of gender identity in analyzing domestic violence cases. To implement this test, we use the ``intimate partner violence'' indicator as the outcome variable of Equation \eqref{EqDV}.

Our second test verifies whether female judges are more likely to make a conviction decision in a domestic violence case and simultaneously flag the case as being related to intimate partner violence. If there is a gender conviction-rate gap in cases flagged as intimate partner violence but not in cases without such a flag, then a stronger identification with the victim may be driving the gender lenience gap found in Section 4. If the gender conviction-rate gap is present regardless of explicit mentions to intimate partner violence, then there must be other forces driving this lenience gap. To implement this test, we interact the ``intimate partner violence'' indicator with the conviction indicator and use this interaction term as the outcome variable of Equation \eqref{EqDV}.

Table \ref{table:Partner} shows the results of these two tests. In Column (1), the outcome is a dummy variable indicating whether the judge's ruling mentioned intimate partner violence. In Column (2), the outcome is a dummy variable indicating whether the judge's ruling mentioned intimate partner violence and reached a conviction decision. In Column (3), the outcome is a dummy variable indicating whether the judge's ruling did not mention intimate partner violence but reached a conviction decision. In Column (4), the outcome is a dummy variable indicating whether the case resulted in a conviction, replicating the results from Table \ref{table: dv_difference} for convenience.

\begin{table}[!htb]
	\caption{Differences by Judge Gender in the Probability of Conviction and the Salience of Intimate Partner Violence}\label{table:Partner}
	\begin{center}
		\begin{tabular}{l c c c c}
			\hline \hline
			& \multirow{2}{*}{Partner} & Convicted and  & Convicted and & \multirow{2}{*}{Convicted} \\
			&  & Partner & Not Partner & \\
			& (1) & (2) & (3) & (4) \\
			\hline
			Female & $0.13^{***}$ & $0.09^{***}$   & $0.01$ & $0.10^{***}$ \\
			& $(0.03)$ & $(0.03)$ & $(0.02)$ &  $(0.03)$   \\
			\hline
			Men's Outcome Average & $0.57$ & $0.26$ & $0.09$ & $0.35$  \\
			District-Quarter FE & $\checkmark$ & $\checkmark$ & $\checkmark$ & $\checkmark$  \\
			Case Controls & $\checkmark$ & $\checkmark$ & $\checkmark$ & $\checkmark$ \\
			Career Controls & $\checkmark$ & $\checkmark$ & $\checkmark$  & $\checkmark$ \\
			Num. obs. & $4,215$ & $4,215$ & $4,215$ & $4,215$       \\
			\hline
			\multicolumn{5}{l}{\footnotesize{$^{***}p<0.01$; $^{**}p<0.05$; $^{*}p<0.1$}}
		\end{tabular}
	\end{center}
	\footnotesize{Notes: The regressions in this table uses the sample of domestic violence crimes only. The outcome in Column (1) is whether the sentence mentioned intimate partner violence. The outcome in Column (2) is whether the sentence mentioned intimate partner violence and reached a conviction. The outcome in Column (3) is whether the case did not mention intimate partner violence and reached a conviction. The outcome in Column (4) is whether the sentence reached a conviction. All regressions control for District-Quarter fixed effects, judge's career covariates (years of experience, career stage, years of experience in the current career stage), and case characteristics (defendant was caught red-handed, defendant was arrested by the sheriff before the trial). Standard errors, presented in parentheses, are clustered at the case ID level.}
\end{table}

We find that female judges are 22.1\% (12.51 p.p.) more likely than male judges to mention the relationship status between victim and defendant in their rulings (Column (1)). This difference can arise as a result of female judges being more willing to write longer, more detailed sentences (Section \ref{SecConsequencesAppealsMechanism}) or being more aware of intimate partner violence due to their gender identity.

Additionally, we find that female judges are 34.1\% (9.00 p.p.) more likely than male judges to convict the defendant and simultaneously flag the case as intimate partner violence (Column (2)). However, female judges are as likely as male judges to convict the defendant while not flagging the case as intimate partner violence (Column (3)). These results suggest that the gender conviction-rate gap is higher in cases where a female judge may identify more strongly with the victim compared to a male judge.

These findings are consistent with existing evidence that, in many contexts, judges make different decisions when they share a group identity with defendants or plaintiffs \citep{kruttschnitt_ages_2009, shayo_judicial_2011, shayo_conflict_2017, cai_judges_2022, chenJudgesFavorTheir2022}. Many of these papers show that these differences intensify as group identity becomes more salient. This intensification may result from a heightened opposition between two or more social groups (``us'' versus ``them''). For example, \citet{shayo_judicial_2011} show that Jewish Israeli judges in small claim courts favor Jewish litigants more intensely when they are more exposed to conflict-related terrorism. Closer to our case, \citet{cai_judges_2022} show that judges analyzing divorce cases in China favor plaintiffs of their own gender more in areas where social norms are more conservative. In our case, we show that the gender conviction-rate gap is more intense in cases where the form of domestic violence {might be considered} more stereotypical.\footnote{Note that most domestic violence cases against women are committed by partners and ex-partners \citep{united_nations_office_on_drugs_and_crime_gender-related_2023}.} {Consequently, the representational account \citep{boyd_untangling_2010} is a possible driving force of the in-group bias documented in Section \ref{SecResults} because female judges tend to protect members of their own group more intensely than male judges.}

{
	As a caveat, we point that these results are only suggestive because there might be alternative explanations for them. For example, intimate partner violence cases may involve more severe types of violence (e.g., repeated abuse) than other types of domestic violence and female judges might react more strongly than male judges to severe types of violence regardless of any gender identity issues. We cannot rule out this mechanism using a strategy similar to Equation (2) because our comparable crimes do not involve intimate partner violence.
}

\newpage

\section{Additional Tables} \label{AppTable}

\setcounter{table}{0}
\renewcommand\thetable{D.\arabic{table}}

\setcounter{figure}{0}
\renewcommand\thefigure{D.\arabic{figure}}

\setcounter{equation}{0}
\renewcommand\theequation{D.\arabic{equation}}

\setcounter{theorem}{0}
\renewcommand\thetheorem{D.\arabic{theorem}}

\setcounter{proposition}{0}
\renewcommand\theproposition{D.\arabic{proposition}}

\setcounter{corollary}{0}
\renewcommand\thecorollary{D.\arabic{corollary}}

\setcounter{assumption}{0}
\renewcommand\theassumption{D.\arabic{assumption}}

\setcounter{definition}{0}
\renewcommand\thedefinition{D.\arabic{definition}}

\setcounter{Lemma}{0}
\renewcommand\theLemma{D.\arabic{Lemma}}

\begin{table}[!htbp]
	\caption{Differences by Judge Gender in the Probability of Conviction Comparing Domestic Violence Cases against Similar Crimes with flexible controls}\label{table: diff_in_diff flexible}
	\begin{center}{
			\begin{tabular}{l c c }
				\hline \hline
				Sample & Misdemeanors & Physical Assault \\
				& (1) & (2) \\
				\hline
				Female & $0.03^{***}$ & $-0.0001$ \\
				& $(0.01)$ & $(0.043)$ \\
				Domestic Violence (DV) & $-0.09^{**}$ & $-0.01$ \\
				& $(0.04)$ & $(0.11)$ \\
				DV $\times$ Female & $0.02$ & $0.09^{*}$ \\
				& $(0.02)$ & $(0.05)$ \\
				\hline
				Men's CR (Non-DV) & $0.44$ & $0.32$ \\
				Men's CR (All) & $0.43$ & $0.34$ \\
				District-Quarter FE & $\checkmark$ & $\checkmark$ \\
				Career Controls & $\checkmark$ & $\checkmark$ \\
				Case Controls & $\checkmark$ & $\checkmark$ \\
				Num. obs. & 48,213 & 5,903 \\
				\hline
				\multicolumn{3}{l}{\footnotesize{$^{***}p<0.01$; $^{**}p<0.05$; $^{*}p<0.1$}}
		\end{tabular}}
	\end{center}
	\footnotesize{Notes: This table presents the estimated results of Equation \eqref{EqDVComparable}. The first column compares differences in male and female judge conviction rates in domestic violence cases to the gender difference in conviction rates for misdemeanors. The second column compares differences in male and female judge conviction rates in domestic violence cases to the gender difference in conviction rates for other physical assault crimes. All regressions control for judge's career covariates (years of experience, career stage, years of experience in the current career stage), case characteristics (defendant was caught red-handed, defendant was arrested by the sheriff before the trial), and their interactions with the domestic violence indicator. CR stands for Conviction Rate and DV stands for Domestic Violence. Standard errors, presented in parentheses, are clustered at the case ID level.}
\end{table}

\phantom{a}

\newpage

\begin{table}[p]
	\caption{Differences by Judge Gender in the Probability of a Defendant Recidivating using the Misdemeanor Sample}\label{TabRecidivismMisdemeanor}
	\begin{center}{
			\begin{tabular}{l c c c c}
				\hline \hline
				\multicolumn{5}{c}{\emph{Panel A: All cases regardless of conviction status}} \\ \hline
				& Any & Excluding DV & Only DV & Murder \\
				& (1) & (2) & (3) & (4) \\
				\hline
				Female & $0.01$ & $0.00$ & $0.004$ & $0.002$ \\
				& $(0.01)$ & $(0.01)$ & $(0.004)$ & $(0.004)$ \\
				DV  & $-0.022$ & $-0.06^{***}$ & $0.04^{***}$ & $0.014^{**}$ \\
				& $(0.015)$ & $(0.01)$ & $(0.01)$ & $(0.005)$ \\
				DV $\times$ Female  & $0.01$ & $0.02$ & $-0.003$ & $-0.002$ \\
				& $(0.03)$ & $(0.02)$ & $(0.013)$ & $(0.010)$ \\ \hline
				\multicolumn{5}{l}{\emph{Male Judges' Average Outcome for different types of cases}} \\
				Non-DV & $0.35$ & $0.32$ & $0.03$ & $0.03$ \\
				All & $0.35$ & $0.31$ & $0.04$ & $0.03$ \\ \hline
				Num. obs.  & $21,322$ & $21,322$ & $21,322$ & $21,322$  \\ \hline
				\multicolumn{5}{c}{\emph{Panel B: Convicted cases only (Outcome-based Test)}} \\ \hline
				& Any & Excluding DV & Only DV & Murder \\
				& (1) & (2) & (3) & (4) \\
				\hline
				Female & $-0.01$ & $-0.005$ & $-0.002$ & $0.00$ \\
				& $(0.02)$ & $(0.022)$ & $(0.008)$ & $(0.01)$ \\
				DV  & $-0.02$ & $-0.08^{***}$ & $0.05^{***}$ & $0.02^{*}$ \\
				& $(0.03)$ & $(0.03)$ & $(0.01)$ & $(0.01)$ \\
				DV $\times$ Female  & $0.03$ & $-0.02$ & $0.048^{*}$ & $-0.02$ \\
				& $(0.05)$ & $(0.05)$ & $(0.029)$ & $(0.02)$ \\ \hline
				\multicolumn{5}{l}{\emph{Male Judges' Average Outcome for different types of cases}} \\
				Non-DV & $0.40$ & $0.36$ & $0.04$ & $0.03$ \\
				All & $0.39$ & $0.35$ & $0.04$ & $0.03$ \\ \hline
				Num. obs.  & $8,613$ & $8,613$ & $8,613$ & $8,613$  \\ \hline
				District-Quarter FE & $\checkmark$ & $\checkmark$ & $\checkmark$ & $\checkmark$ \\
				Career Controls & $\checkmark$ & $\checkmark$ & $\checkmark$ & $\checkmark$ \\
				Case Controls & $\checkmark$ & $\checkmark$ & $\checkmark$ & $\checkmark$ \\
				\hline
				\multicolumn{5}{l}{\footnotesize{$^{***}p<0.01$; $^{**}p<0.05$; $^{*}p<0.1$}}
		\end{tabular}}
	\end{center}
	\footnotesize{Notes: The regressions (Equation \eqref{EqDVComparable}) in this table use the sample of misdemeanor cases. Panel A uses all case-defendant pairs regardless of the case's conviction status. Panel B uses only case-defendant pairs whose defendants were convicted, implementing an outcome-based test. The outcome in Column (1) indicates whether the defendant recidivated within 2 years of the date of the final ruling. The outcome in Column (2) indicates whether the defendant recidivated within 2 years of the date of the final ruling and the second offense was not a domestic violence case. The outcome in Column (3) indicates whether the defendant recidivated within 2 years of the date of the final ruling and the second offense was a domestic violence case. The outcome in Column (4) indicates whether the defendant recidivated within 2 years of the date of the final ruling and the second offense was a crime against someone's life (e.g., manslaughter, murder, attempted murder). All regressions control for District-Quarter fixed effects, judge's career covariates (years of experience, career stage, years of experience in the current career stage), and case characteristics (defendant was caught red-handed, defendant was arrested by the sheriff before the trial). Standard errors, presented in parentheses, are clustered at the case ID level. The sample size in the recidivism regressions is smaller than in the other regressions because we do not use the last two years of data to avoid right-censoring problems.}
\end{table}

\newpage

\begin{table}[p]
	\caption{Differences by Judge Gender in the Probability of a Defendant Recidivating using the Physical Assault Sample}\label{TabRecidivismPA}
	\begin{center}{
			\begin{tabular}{l c c c c}
				\hline \hline
				\multicolumn{5}{c}{\emph{Panel A: All cases regardless of conviction status}} \\ \hline
				& Any & Excluding DV & Only DV & Murder \\
				& (1) & (2) & (3) & (4) \\
				\hline
				Female & $-0.03$ & $-0.07$ & $0.05$ & $0.01$ \\
				& $(0.06)$ & $(0.06)$ & $(0.03)$ & $(0.03)$ \\
				DV  & $0.01$ & $-0.02$ & $0.03$ & $-0.002$ \\
				& $(0.04)$ & $(0.04)$ & $(0.02)$ & $(0.019)$ \\
				DV $\times$ Female  & $0.05$ & $0.08$ & $-0.03$ & $-0.003$ \\
				& $(0.06)$ & $(0.06)$ & $(0.03)$ & $(0.028)$ \\ \hline
				\multicolumn{5}{l}{\emph{Male Judges' Average Outcome for different types of cases}} \\
				Non-DV & $0.25$ & $0.22$ & $0.04$ & $0.03$ \\
				All & $0.30$ & $0.24$ & $0.06$ & $0.04$ \\ \hline
				Num. obs.  & $3,467$ & $3,467$ & $3,467$ & $3,467$  \\ \hline
				\multicolumn{5}{c}{\emph{Panel B: Convicted cases only (Outcome-based Test)}} \\ \hline
				& Any & Excluding DV & Only DV & Murder \\
				& (1) & (2) & (3) & (4) \\
				\hline
				Female & $-0.09$ & $-0.08$ & $-0.01$ & $0.02$ \\
				& $(0.19)$ & $(0.19)$ & $(0.10)$ & $(0.03)$ \\
				DV  & $0.04$ & $-0.02$ & $0.05$ & $0.01$ \\
				& $(0.11)$ & $(0.10)$ & $(0.07)$ & $(0.02)$ \\
				DV $\times$ Female  & $0.12$ & $0.08$ & $0.04$ & $-0.02$ \\
				& $(0.19)$ & $(0.18)$ & $(0.11)$ & $(0.03)$ \\ \hline
				\multicolumn{5}{l}{\emph{Male Judges' Average Outcome for different types of cases}} \\
				Non-DV & $0.28$ & $0.28$ & $0.03$ & $0.01$ \\
				All & $0.34$ & $0.27$ & $0.07$ & $0.04$ \\ \hline
				Num. obs.  & $1,132$ & $1,132$ & $1,132$ & $1,132$  \\ \hline
				District-Quarter FE & $\checkmark$ & $\checkmark$ & $\checkmark$ & $\checkmark$ \\
				Career Controls & $\checkmark$ & $\checkmark$ & $\checkmark$ & $\checkmark$ \\
				Case Controls & $\checkmark$ & $\checkmark$ & $\checkmark$ & $\checkmark$ \\
				\hline
				\multicolumn{5}{l}{\footnotesize{$^{***}p<0.01$; $^{**}p<0.05$; $^{*}p<0.1$}}
		\end{tabular}}
	\end{center}
	\footnotesize{Notes: The regressions (Equation \eqref{EqDVComparable}) in this table use the sample of physical assault cases. Panel A uses all case-defendant pairs regardless of the case's conviction status. Panel B uses only case-defendant pairs whose defendants were convicted, implementing an outcome-based test. The outcome in Column (1) indicates whether the defendant recidivated within 2 years of the date of the final ruling. The outcome in Column (2) indicates whether the defendant recidivated within 2 years of the date of the final ruling and the second offense was not a domestic violence case. The outcome in Column (3) indicates whether the defendant recidivated within 2 years of the date of the final ruling and the second offense was a domestic violence case. The outcome in Column (4) indicates whether the defendant recidivated within 2 years of the date of the final ruling and the second offense was a crime against someone's life (e.g., manslaughter, murder, attempted murder). All regressions control for District-Quarter fixed effects, judge's career covariates (years of experience, career stage, years of experience in the current career stage), and case characteristics (defendant was caught red-handed, defendant was arrested by the sheriff before the trial). Standard errors, presented in parentheses, are clustered at the case ID level. The sample size in the recidivism regressions is smaller than in the other regressions because we do not use the last two years of data to avoid right-censoring problems.}
\end{table}

\begin{landscape}
	\begin{table}[!htb]
		\caption{Differences by Judge Gender in the Probability of Non-Prosecution Agreement, Dismissal and Acquittal}\label{TabOtherSentences}
		\begin{center}
			\begin{tabular}{l c c c c c c c c c c c}
				\hline \hline
				& \multicolumn{3}{c}{Sample:}& & \multicolumn{3}{c}{Sample:} & & \multicolumn{3}{c}{Sample:} \\
				& \multicolumn{3}{c}{Domestic Violence }& & \multicolumn{3}{c}{Misdemeanors} & & \multicolumn{3}{c}{Physical Assault } \\ \cline{2-4} \cline{6-8} \cline{10-12}
				& NPA & Dismissal & Acquittal & & NPA & Dismissal & Acquittal & & NPA & Dismissal & Acquittal \\
				& (1) & (2) & (3) & & (4) & (5) & (6) & & (7) & (8) & (9)  \\
				\hline
				Fem & $-0.002$ & $-0.003$ & $-0.02$ & & $0.012^{***}$ & $0.01$ & $-0.03^{***}$ & & $-0.01$ & $0.02$ & $0.02$ \\
				& $(0.006)$ & $(0.011)$ & $(0.04)$ & & $(0.003)$ & $(0.01)$ & $(0.01)$ & & $(0.02)$ & $(0.03)$ & $(0.05)$    \\
				DV  &  &  &  & & $-0.019^{***}$ & $-0.10^{***}$ & $0.18^{***}$ & & $ -0.05^{***}$ & $-0.13^{***}$ & $0.12^{***}$ \\
				&  &  &  & & $(0.003)$ & $(0.01)$ & $(0.01)$ & & $(0.01)$ & $(0.02)$ & $(0.03)$    \\
				DV $\times$ Fem  &  &  &  & & $-0.022^{***}$ & $-0.002$ & $-0.0001$ & & $0.00$ & $-0.01$ & $-0.04$ \\
				&  &  &  & & $(0.005)$ & $(0.009)$ & $(0.0193)$ & & $(0.02)$ & $(0.03)$ & $(0.05)$    \\
				\hline
				Men's CR (Non-DV) & $-$ & $-$ & $-$ & & $0.04$ & $0.13$ & $0.30$ & & $0.06$ & $0.17$ & $0.34$ \\
				Men's CR (All) & $0.01$ & $0.04$ & $0.49$ & & $0.04$ & $0.12$ & $0.32$ & & $0.02$ & $0.07$ & $0.45$ \\
				District-Quarter FE & $\checkmark$ & $\checkmark$ & $\checkmark$ & & $\checkmark$ & $\checkmark$ & $\checkmark$ & & $\checkmark$ & $\checkmark$ & $\checkmark$ \\
				Career Controls & $\checkmark$ & $\checkmark$ & $\checkmark$ & & $\checkmark$ & $\checkmark$ & $\checkmark$ & & $\checkmark$ & $\checkmark$ & $\checkmark$ \\
				Case Controls & $\checkmark$ & $\checkmark$ & $\checkmark$ & & $\checkmark$ & $\checkmark$ & $\checkmark$ & & $\checkmark$ & $\checkmark$ & $\checkmark$ \\
				Num. obs.   & $4,215$ & $4,215$ & $4,215$ & & $48,213$ & $48,213$ & $48,213$ & & $5,903$ & $5,903$ & $5,903$ \\
				\hline
				\multicolumn{12}{l}{\footnotesize{$^{***}p<0.01$; $^{**}p<0.05$; $^{*}p<0.1$}}
			\end{tabular}
		\end{center}
		\footnotesize{Notes: This table shows the results of Equations \eqref{EqDV} and \eqref{EqDVComparable} using three outcomes. In Columns (1), (4) and (7), the outcome indicates whether the case reached a non-prosecution agreement. In Columns (2), (5) and (8), the outcome indicates whether the judge decided to dismiss the case. In Columns (3), (6) and (9), the outcome indicates whether the judge acquitted the defendant. The first three columns use the sample of domestic violence cases (Equation \eqref{EqDV}). Columns (4)-(6) use the sample of misdemeanors, while Columns (7)-(9) use the sample of physical assault crimes (Equation \eqref{EqDVComparable}). DV stands for Domestic Violence. All regressions control for District-Quarter fixed effects (FE), judge's career covariates (years of experience, career stage, years of experience in the current career stage), and case characteristics (defendant was caught red-handed, defendant was arrested by the sheriff before the trial). Standard errors, presented in parentheses, are clustered at the case ID level.}
	\end{table}
\end{landscape}

\newpage

\section{Heterogeneity around the Me Too Movement} \label{AppMeToo}

\setcounter{table}{0}
\renewcommand\thetable{E.\arabic{table}}

\setcounter{figure}{0}
\renewcommand\thefigure{E.\arabic{figure}}

\setcounter{equation}{0}
\renewcommand\theequation{E.\arabic{equation}}

\setcounter{theorem}{0}
\renewcommand\thetheorem{E.\arabic{theorem}}

\setcounter{proposition}{0}
\renewcommand\theproposition{E.\arabic{proposition}}

\setcounter{corollary}{0}
\renewcommand\thecorollary{E.\arabic{corollary}}

\setcounter{assumption}{0}
\renewcommand\theassumption{E.\arabic{assumption}}

\setcounter{definition}{0}
\renewcommand\thedefinition{E.\arabic{definition}}

\setcounter{Lemma}{0}
\renewcommand\theLemma{E.\arabic{Lemma}}

{In this appendix, we evaluate whether the gender gap in conviction rates varies over time.} To do so, we analyze whether the $\beta$ coefficient in Equation \eqref{EqDV} and the $\beta_{3}$ coefficient in Equation \eqref{EqDVComparable} vary as a function of time. Specifically, we create two subsamples and estimate Equations \eqref{EqDV} and \eqref{EqDVComparable} for each one of them. The first subsample contains the cases whose trial judges' rulings were written before the Me Too Movement (October 1\textsuperscript{st}, 2017), while the second subsample contains the cases whose trial judges' rulings were written after the Me Too Movement.

We chose to analyze the heterogeneity before and after the Me Too Movement because judges' behavior might have changed around this event. In particular, the public repercussion of the Me Too Movement may have increased male judges' awareness of violence against women. As a consequence, male judges' analyses of domestic violence cases may have become more similar to those of their female peers. Empirically, this increased similarity between male and female judges will decrease the $\beta$ coefficient in Equation \eqref{EqDV} and the $\beta_{3}$ coefficient in Equation \eqref{EqDVComparable}.

Table \ref{TabMeToo} shows estimates of Equations \eqref{EqDV} and \eqref{EqDVComparable} before and after the Me Too Movement.\footnote{We emphasize that this before-and-after comparison is a heterogeneity exercise only, since our identification strategy is not appropriate for estimating the causal effect of the Me Too Movement on judges' behavior.} Panel A uses only case-defendant pairs whose final ruling was written before the start of the Me Too Movement (October 1\textsuperscript{st}, 2017), while Panel B uses only case-defendant pairs whose final ruling was written after the start of the Me Too Movement. Column (1) uses the sample of domestic violence cases (Equation \eqref{EqDV}), while Columns (2) and (3) use the samples of misdemeanor cases and physical assault cases (Equation \eqref{EqDVComparable}). All regressions control for district-quarter fixed effects, judge's career covariates (years of experience, career stage, years of experience in the current career stage), and case characteristics (defendant was caught red-handed, defendant was arrested by the sheriff before the trial).

\begin{table}[!htbp]
	\caption{Differences by Judge Gender in the Probability of Conviction before and after the Me Too Movement}\label{TabMeToo}
	\begin{center}{
			\begin{tabular}{l c c c}
				\hline \hline
				\multicolumn{4}{c}{\emph{Panel A: Before the Me Too Movement}} \\ \hline
				Sample & DV Only & Misdemeanor & Physical Assault \\
				& (1) & (2) & (3)\\
				\hline
				Female & $0.18^{***}$ & $0.07^{***}$ & $0.06$ \\
				& $(0.04)$ & $(0.01)$ & $(0.07)$ \\
				DV   & & $-0.09^{***}$ & $0.04$ \\
				& & $(0.02)$ & $(0.04)$ \\
				DV $\times$ Female & & $0.05^{*}$ & $0.09$ \\
				& & $(0.03)$ & $(0.07)$ \\ \hline
				\multicolumn{4}{l}{\emph{Male Judges' Average Outcome for different types of cases}} \\
				Non-DV & $-$ & $0.42$ & $0.30$ \\
				All & $0.33$ & $0.41$ & $0.32$ \\ \hline
				Num. obs.  & $2,501$ & $19,449$ & $3,227$ \\ \hline
				\multicolumn{4}{c}{\emph{Panel B: After the Me Too Movement}} \\ \hline
				Sample & DV Only & Misdemeanor & Physical Assault \\
				& (1) & (2) & (3)\\
				\hline
				Female & $0.05$ & $0.01$ & $-0.02$ \\
				& $(0.06)$ & $(0.01)$ & $(0.07)$ \\
				DV   & & $-0.06^{***}$ & $0.02$ \\
				& & $(0.02)$ & $(0.05)$ \\
				DV $\times$ Female & & $-0.02$ & $0.08$ \\
				& & $(0.03)$ & $(0.08)$ \\ \hline
				\multicolumn{4}{l}{\emph{Male Judges' Average Outcome for different types of cases}} \\
				Non-DV & $-$ & $0.46$ & $0.34$ \\
				All & $0.38$ & $0.45$ & $0.37$ \\ \hline
				Num. obs.  & $1,714$ & $28,764$ & $2,676$ \\ \hline
				District-Quarter FE & $\checkmark$ & $\checkmark$ & $\checkmark$\\
				Career Controls & $\checkmark$ & $\checkmark$ & $\checkmark$\\
				Case Controls & $\checkmark$ & $\checkmark$ & $\checkmark$\\
				\hline
				\multicolumn{4}{l}{\footnotesize{$^{***}p<0.01$; $^{**}p<0.05$; $^{*}p<0.1$}}
		\end{tabular}}
	\end{center}
	\footnotesize{Notes: This table shows the results of Equations \eqref{EqDV} and \eqref{EqDVComparable} for two sub-samples. Panel A uses only case-defendant pairs whose final ruling was written before the start of the Me Too Movement (October 1\textsuperscript{st}, 2017), while Panel B uses only case-defendant pairs whose final ruling was written after the start of the Me Too Movement. Column (1) uses the sample of domestic violence cases (Equation \eqref{EqDV}), while Columns (2) and (3) use the samples of misdemeanor cases and physical assault cases (Equation \eqref{EqDVComparable}). DV stands for Domestic Violence. All regressions control for District-Quarter fixed effects, judge's career covariates (years of experience, career stage, years of experience in the current career stage), and case characteristics (defendant was caught red-handed, defendant was arrested by the sheriff before the trial). Standard errors, presented in parentheses, are clustered at the case ID level.}
\end{table}

We find that the results in Panel A are similar to the results in Tables \ref{table: dv_difference} and \ref{table: diff_in_diff}. This finding suggests that, before the Me Too Movement, there was a significant gender conviction rate gap (``Female'' coefficient in Column (1)) and a significant in-group bias (``DV $\times$ Female'' coefficient in Column (2)). Despite not being significant, the interaction coefficient in Column (3) suggests that, before the Me Too Movement, the gender conviction rate gap for domestic violence cases is 143.35\% larger than the same gap for physical assault cases.

However, these coefficients are closer to zero and non-significant in Panel B. These results suggest that, after the Me Too Movement, the gender conviction rate gap and the in-group bias became smaller.

\end{document}